\documentclass[11pt, a4paper]{article}

\usepackage{amsmath}
\usepackage{amsthm}
\usepackage{natbib}
\usepackage{graphicx}
\usepackage{booktabs}

\usepackage{pdfpages}
\usepackage[left= 2.5cm, right = 2.5cm, top = 3cm, bottom = 3cm]{geometry}

\usepackage{color}

\newtheoremstyle{example}
{3pt} 
{3pt} 
{} 
{0\parindent} 
{\bf}
 {\scshape} 
{:} 
{.5em} 
{} 
\newtheoremstyle{theorem}
{3pt} 
{3pt} 
{\em} 
{0\parindent} 
{\bf}
 {\scshape} 
{:} 
{.5em} 
{} 
\theoremstyle{example}

\theoremstyle{theorem}

\def\T{{ \mathrm{\scriptscriptstyle T} }}
\DeclareMathOperator{\Tr}{tr}

\begin{document}
\title{Median bias reduction in random-effects meta-analysis and meta-regression}

\author{
S.~Kyriakou \\
Department of Statistical Science \\
University College London \\
London, WC1E 6BT, UK \\
\texttt{sophia.kyriakou.14@ucl.ac.uk}
\bigskip \\
I.~Kosmidis \\
Department of Statistics \\
University of Warwick \\
Coventry, CV4 7AL, UK \smallskip \\
The Alan Turing Institute \\
96 Euston Road, London NW1 2DB, UK \\
\texttt{Ioannis.Kosmidis@warwick.ac.uk}
\bigskip \\
N.~Sartori \\
Department of Statistical Sciences \\
University of Padova \\
Via Cesare Battisti, 35121 Padova, Italy \\
\texttt{sartori@stat.unipd.it}
}

\maketitle

\begin{abstract}
 The reduction of the mean or median bias of the maximum likelihood estimator in regular parametric models can be
achieved through the additive adjustment of the score equations. In this paper, we derive the adjusted score equations
for median bias reduction in random-effects meta-analysis and meta-regression models and derive efficient estimation
algorithms. The median bias-reducing adjusted score functions are found to be the derivatives of a penalised likelihood.
The penalised likelihood is used to form a penalised likelihood ratio statistic which has known limiting distribution and
can be used for carrying out hypothesis tests or for constructing confidence intervals for either the fixed-effect
parameters or the variance component. Simulation studies and real data applications are used to assess the
performance of estimation and inference based on the median bias-reducing penalised likelihood and compare it to
recently proposed alternatives. The results provide evidence on the effectiveness of median bias reduction in improving
estimation and likelihood-based inference.

\end{abstract}

\section{Introduction}

Meta-analysis is a core tool for synthesizing the results from
independent studies investigating a common effect of interest. One of
the main challenges when combining results from multiple studies is
the variability or heterogeneity in the design and the methods
employed in each study. Accounting for and quantifying that
heterogeneity is critical when drawing inferences about the common
effect. In this direction, \citet{dersimonian1986meta} introduced the
random-effects meta-analysis model, which expresses the heterogeneity
between studies in terms of a variance component that can be estimated
through standard estimation techniques.

Nevertheless, there is ample evidence that frequentist inference based
on random-effects meta-analysis can be problematic in the usual
meta-analytic scenario where the number of studies is small or
moderate. Specifically, the estimation of the heterogeneity parameter
can be highly imprecise, which in turn results in misleading
conclusions \citep{van2002advanced,guolo2015random,kosmidis2017improving}.
Examples of recently proposed methods that attempt to improve
inference are the resampling \citep{jackson2009re} and double resampling
approaches\citep{zeng2015random}, and the mean bias-reducing penalized
likelihood (BRPL) approach in
\citet{kosmidis2017improving}. Specifically,
\citet{kosmidis2017improving} show that maximization of the BRPL
results in an estimator of the heterogeneity parameter that has
notably smaller bias than maximum likelihood (ML) with small loss in
efficiency, and illustrate that BRPL-based inference outperforms its
competitors in terms of inferential performance.

\citet{pagui2016median} show that under suitable conditions
third-order median unbiased estimators can be obtained by the solution
of a suitably adjusted score equation. The components of such median bias-reduced
estimators have, to third-order, the same probability of over- and under-estimating the true parameter. A key property of these estimators, not shared with the mean bias-reduced ones,
is that any monotone component-wise transformation of the estimators
results automatically in median bias-reduced estimators of the
transformed parameters \citep{pagui2016median}. Such equivariance
property can be useful in the context of random-effects meta-analysis
where the Fisher information and, hence, the asymptotic variances of
various likelihood-based estimators depend only on the heterogeneity
parameter.

In this paper, we derive the median bias-reducing adjusted score
functions for random-effects meta-analysis and meta-regression. The
adjusted score functions are found to correspond to a median BRPL,
whose logarithm differs from the logarithm of the mean BRPL in
\citet{kosmidis2017improving} by a simple additive term that depends
on the heterogeneity parameter. Since the adjustments to the score
function for mean and median bias reduction are both of order $O(1)$,
the same arguments as in \cite{kosmidis2017improving}
are used to obtain a median BRPL ratio statistic with known asymptotic
null distribution that can be used for carrying out hypothesis tests
and constructing confidence regions or intervals for either the fixed-effect or the heterogeneity parameter. Simulation studies and real data applications are used to assess the
performance of estimation and inference based on the median BRPL, and
compare it to recently proposed alternatives, including the mean
BRPL. The results provide evidence on the effectiveness of median bias
reduction in improving estimation and likelihood-based inference.


\section{Cocoa intake and blood pressure reduction data}
\label{section:cocoa}

\begin{figure}
\centering
\includegraphics[scale=0.55]{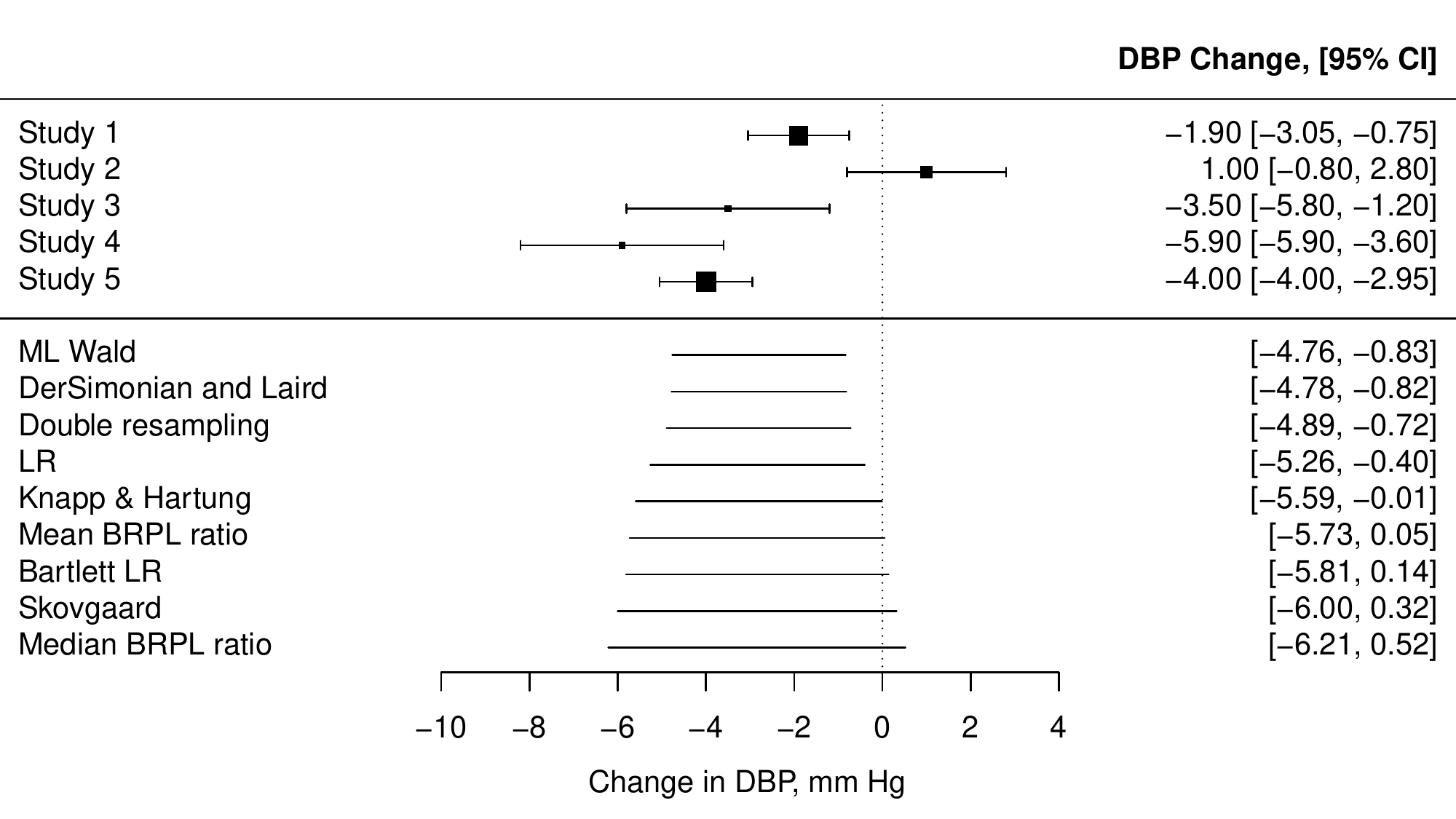}
\caption{Forest plot of cocoa data \citep{bellio2016integrated,
    taubert2007effect}. The outcomes from the five studies are
  reported in terms of the diastolic blood pressure (DBP) difference
  after two weeks of cocoa consumption. A negative change in DBP
  indicates favorable hypotensive cocoa actions. Squares represent the
  mean effect estimate for each study; the size of the square reflects
  the weight that the corresponding study exerts in the meta-analysis
  calculated as the within-study's inverse variance. Horizontal line
  segments represent $95\%$ Wald-type confidence intervals (CI) of the
  effect estimate of individual studies. In the bottom panel of the
  plot horizontal line segments represent the corresponding $95\%$
  confidence interval as computed based on the Wald statistic using
  the ML estimates (ML Wald), the DerSimonian and Laird approach
  \citep{dersimonian1986meta}, double resampling
  \citep{zeng2015random}, the LR statistic, the
  \citet{knapp2003improved} method, the mean BRPL ratio statistic
  \cite{kosmidis2017improving}, the Bartlett-corrected LR statistic
  (Bartlett LR) \citep{huizenga2011testing}, the Skovgaard's
  statistic, and the median BRPL ratio statistic. The confidence
  intervals are ordered according to their length. The estimate of
  $\beta$ has not been reported, as is commonly done in forest plots,
  because some of the methods considered (e.g.~Skovgaard,
  Bartlett-corrected LR, and double resampling) are designed to
  produce directly $p$-values and/or confidence intervals and do not
  directly correspond to an estimation method.}
\label{forest:cocoa}
\end{figure}

Consider the setting in \citet{bellio2016integrated} who carry out a meta-analysis of five randomized controlled trials from \citet{taubert2007effect} on the efficacy of two weeks of cocoa consumption on lowering diastolic blood
pressure. Figure~\ref{forest:cocoa} is a forest plot with the
estimated mean difference in diastolic blood pressure before and after cocoa intake from each study, and the associated $95\%$ Wald-type confidence intervals. Four out of the five studies reported a reduction of diastolic blood pressure from cocoa intake.

The random-effects meta-analysis model is used to synthesize the evidence from the five studies. In particular, let $Y_{i}$ be the random variable representing the mean difference in the diastolic blood pressure after two weeks of cocoa intake in the $i$th study. We assume that $Y_{1}, \ldots, Y_{5}$ are independent random variables where $Y_i$ has a Normal distribution with mean the overall effect $\beta$ and variance $\hat\sigma^2_i + \psi$, with $\hat\sigma^2_i$ the estimated standard error of the effect from the $i$th study and $\psi$ the heterogeneity parameter.

The forest plot in Figure~\ref{forest:cocoa} depicts nominally $95\%$
confidence intervals for $\beta$ using various alternative methods. As
is apparent, the conclusions when testing the hypothesis $\beta = 0$
can vary depending on the method used. More specifically, the Wald test using the ML estimates,
the DerSimonian and Laird method \citep{dersimonian1986meta}, double
resampling \citep{zeng2015random}, and the likelihood ratio (LR) test
give evidence that there is a relationship between cocoa consumption
and diastolic blood pressure, with $p$-values $0.005$, $0.006$,
$0.016$, $0.030$, respectively. On the other hand, Knapp and Hartung's
method \citep{knapp2003improved}, the mean BRPL
ratio \citep{kosmidis2017improving}, the Bartlett-corrected
LR \citep{huizenga2011testing}, and Skovgaard's test suggest that the
evidence that cocoa consumption affects diastolic blood pressure is
weaker, with $p$-values $0.050$, $0.053$, $0.058$, $0.067$,
respectively.

\section{Random-effects meta-regression model}
\label{section:REmodel}
Let $y_i$ and $\hat{\sigma}_{i}^2$ denote the estimate of the effect
from the $i$th study $(i = 1, \ldots, K)$ and the associated
within-study variance, respectively, and
$x_i = (x_{i1},\ldots,x_{ip})^{\T}$ denote a $p$-vector of
study-specific covariates that can be used to account for the
heterogeneity across studies.

The within-study variances $\hat{\sigma}_{i}^2$ are usually assumed to
be estimated well-enough to be considered as known and equal to the
values reported in each study. Then the observations $y_1,\ldots,y_K$
are assumed to be realizations of the random variables
$Y_1,\ldots,Y_K$, which are independent conditionally on independent
random effects $U_1, \ldots, U_K$. The conditional distribution of
$Y_i$ given $U_i = u_i$ is
$N(u_i + x^{\T}_i\beta, \hat{\sigma}_{i}^2)$, where $\beta$ is an
unknown $p$-dimensional vector of fixed effects. The random effects
$U_1,\ldots,U_K$ are typically assumed to be independent with $U_i$
having a $N(0, \psi)$ distribution, where $\psi$ is a parameter that
attempts to capture the unexplained between-study heterogeneity. In
matrix notation, the random-effects meta-regression model has
\begin{equation}
Y = X \beta + U + \epsilon\,,
\label{eq:metamodel}
\end{equation}
where $Y=(Y_1,\ldots,Y_K)^{\T}$, $X$ is the $K\times p$ model matrix
with $x_i^{\T}$ in its $i$th row, and
$\epsilon=(\epsilon_1,\ldots,\epsilon_K)^{\T}$ is a vector of
independent errors each with a $N(0,\hat{\sigma}_{i}^2)$ distribution
and independent of $U=(U_1,\ldots,U_K)^{\T}$. Under this
specification, the marginal distribution of $Y$ is multivariate normal
with mean $X\beta$ and variance-covariance matrix
$\hat{\Sigma} + \psi I_{K}$, where $I_{K}$ is the $K \times K$
identity matrix and
$\hat{\Sigma} =
\text{diag}(\hat{\sigma}_{1}^2,\ldots,\hat{\sigma}_{K}^2)$. The
random-effects meta-analysis results as a special case of
meta-regression, by setting $X$ to be a column of ones.

The log-likelihood function for $\theta=(\beta^{\T},\psi)^{\T}$ is $l(\theta) = \{\log |W(\psi)| - R(\beta)^{\T}W(\psi)R(\beta)\}/2\,$, where $|W(\psi)|$ denotes the determinant of $W(\psi) = (\hat{\Sigma} + \psi I_{K})^{-1}$ and $R(\beta)=y-X\beta$. The gradient of the log-likelihood (score function) is
\begin{equation}
s(\theta) = \begin{pmatrix}
X^{\T}W(\psi)R(\beta) \\
\frac{1}{2}\{R(\beta)^{\T}W(\psi)^2R(\beta) - \Tr[W(\psi)]\}
\end{pmatrix}\,
\label{eq:metascore}
\end{equation}
and the ML estimator $\hat{\theta}=(\hat{\beta}^{\T},\hat{\psi})^{\T}$
is obtained as the solution of $s(\theta) = 0_{p + 1}$, where $0_p$
denotes a $p$-dimensional vector of zeros.

\section{Median bias reduction}
\label{section:medianBR}
\subsection{The method}

A popular method for reducing the mean bias of ML estimates in regular statistical models is through the adjustment of the score equation \citep{firth1993bias,kosmidis2009bias}. \citet{pagui2016median} propose an extension of the adjusted score equation approach which can be used to obtain median bias-reduced estimators. Specifically, under the model, the new estimator has a distribution with median closer to the ``true" parameter value than the ML estimator. \citet{pagui2016median} consider the median as a centering index for the score, and the adjusted score function for median bias reduction then results by subtracting from the score its approximate median, obtained using a Cornish-Fisher asymptotic expansion.

Let $j(\theta)= -\partial^2 l(\theta)/\partial \theta\partial
\theta^{\T}$ be the observed information matrix (see, Appendix for its
expression), and $i(\theta)$ be the expected information matrix
\begin{equation}
i(\theta) =E_{\theta}(j(\theta)) =  \begin{pmatrix}
X^{\T}W(\psi)X & 0_p \\
0^{\T}_p & \frac{1}{2}\Tr[W(\psi)^2]
\end{pmatrix}\,,
\label{eq:metaregexpectedinfo}
\end{equation}
with $t$th column $i_{t}(\theta)$. Let also $i^{t}(\theta)$ and
$i^{tt}(\theta)$ be the $t$th column and the $t$th diagonal element of
$\{i(\theta)\}^{-1}$, with $t \in
\{1,\ldots,p+1\}$. \citet{pagui2016median} show that a median
bias-reduced estimator $\hat{\theta}^{\dagger}$ can be obtained by
solving an adjusted score equation of the form
$s^{\dagger}(\theta) = s(\theta) + A^{\dagger}(\theta) = 0\,$, where
the additive adjustment to $A^{\dagger}(\theta)$ is $O(1)$, in the
sense that $A^{\dagger}(\theta)$ is bounded in absolute value by a
fixed constant after a sufficiently large value of $K$. The median
bias-reducing adjustment $A^{\dagger}(\theta)$ has $t$th element
\begin{equation}
A^{\dagger}_{t}(\theta) = \frac{1}{2}\Tr\left[\{i(\theta)\}^{-1}(P_{t}(\theta) + Q_{t}(\theta))\right] - \{i_{t}(\theta)\}^{\T}K^{\dagger}(\theta)\,.
\label{eq:medianBRadj}
\end{equation}
The quantities
$P_{t}(\theta) = E_{\theta}[s(\theta)s^{\T}(\theta)s_{t}(\theta)]$ and
$Q_{t}(\theta)=E_{\theta}[-j(\theta)s_{t}(\theta)]$
in~(\ref{eq:medianBRadj}) are those introduced by \citet{kosmidis2009bias} for mean bias-reduction, and
$K^{\dagger}(\theta)$ is a $(p+1)$-vector with $t$th element $K^{\dagger}_{t}(\theta) = \{i^{t}(\theta)\}^{\T}K_{t}(\theta)$, where $K_{t}(\theta)$ is another $(p+1)$-vector with $u$th element
\[
  K_{tu}(\theta) = \Tr\left[\frac{i^{t}(\theta)\{i^{t}(\theta)\}^{\T}}{i^{tt}(\theta)}\left(\frac{1}{3}P_{u}(\theta)
      + \frac{1}{2}Q_{u}(\theta)\right)\right] \, .
\]

In the context of meta-regression values of $t$ and $u$ in $\{1,\ldots,p\}$ correspond to the elements of parameter $\beta$, and $t,u = p+1$ correspond to parameter $\psi$. Given that $A^{\dagger}(\theta)$ is of order $O(1)$, $\hat{\theta}^{\dagger}$ has the same asymptotic distribution as $\hat{\theta}$ \citep{pagui2016median}, i.e.~multivariate normal with mean $\theta$ and variance-covariance matrix $\{i(\theta)\}^{-1}$, which can be consistently estimated with $\{i(\hat{\theta}^{\dagger})\}^{-1}$.

After some algebra (see Appendix for details) the median bias-reducing adjustment for the random-effects meta-analysis and meta-regression models has the form
\begin{equation}
A^{\dagger}(\theta)  = \begin{pmatrix}
0_p \\
\frac{1}{2}\Tr[W(\psi)H(\psi)] + \frac{1}{3}\frac{\Tr[W(\psi)^3]}{\Tr[W(\psi)^2]}
\end{pmatrix}\,,
\label{eq:metaregM1}
\end{equation}
where $H(\psi) = X(X^{\T}W(\psi)X)^{-1}X^{\T}W(\psi)$. Substituting
(\ref{eq:metaregM1}) in the expression for $s^{\dagger}(\theta)$ gives
that the median bias-reducing adjusted score functions for $\beta$ and
$\psi$ are $s^{\dagger}_{\beta}(\theta) = s_{\beta}(\theta)$ and
\begin{equation*}
s^{\dagger}_{\psi}(\theta) = s_{\psi}(\theta) + \frac{1}{2}\Tr[W(\psi)H(\psi)] + \frac{1}{3}\frac{\Tr[W(\psi)^3]}{\Tr[W(\psi)^2]} \,,
\end{equation*}
respectively.

\subsection{Computation of median bias-reduced estimator}
\label{section:algorithm}

A direct approach for computing the estimator
$\hat{\theta}^{\dagger} =
(\hat{\beta}^{\dagger\T},\hat{\psi}^{\dagger})^{\T}$ is through a
modification of the two-step iterative process in
\citet{kosmidis2017improving}. At the $j$th iteration
($j = 1,2,\ldots$)
\begin{enumerate}
\item calculate $\beta^{(j)}$ by weighted least squares as $\beta^{(j)} = (X^{\T}W(\psi^{(j-1)})X)^{-1}X^{\T}W(\psi^{(j-1)})y\,$
\item\label{step:root} solve $s^{\dagger}_{\psi}(\theta^{(j)}(\psi)) = 0$ with respect to $\psi$, where $\theta^{(j)}(\psi)=(\beta^{(j)\T},\psi)^{\T}$.
\end{enumerate}
In the above steps, $\beta^{(j)}$ is the candidate value for
$\hat{\beta}^{\dagger}$ at the $j$th iteration and $\psi^{(j-1)}$ is
the candidate value for $\hat{\psi}^{\dagger}$ at the $(j-1)$th
iteration. The equation in step 2 is solved numerically, by searching
for the root of the function $s^{\dagger}_{\psi}(\beta^{(j)},\psi)$ in
a predefined positive interval. For the computations in this
manuscript we use the \citet{dersimonian1986meta} estimate of $\psi$
as starting value $\psi^{(0)}$. The iterative process is then repeated
until the components of the score function $s^{\dagger}(\theta)$ are
all less than $\epsilon = 1 \times 10^{-6}$ in absolute value at the
current estimates.

\subsection{Median bias-reducing penalized likelihood}

Although it is not generally true that $s^{\dagger}(\theta)$ is the
gradient of a suitable penalized log-likelihood, in this case
$s^{\dagger}(\theta)$ is the gradient of the median BRPL
\begin{equation}
l^{\dagger}(\theta) = l(\theta) - \frac{1}{2}\log |X^{\T}W(\psi)X| - \frac{1}{6}\log [\Tr(W(\psi)^2)]\,.
\label{eq:regpenalizedlik_median}
\end{equation}
Hence, $\hat{\theta}^{\dagger}$ is also the maximum median BRPL estimator. The median BRPL in
(\ref{eq:regpenalizedlik_median}) differs from the mean BRPL derived
in \citet{kosmidis2017improving} by the term
$-\log [\Tr(W(\psi)^2)]/6$.

An advantage of the median BRPL estimators over mean BRPL ones is that
the former are equivariant under monotone component-wise parameter
transformations \citep{pagui2016median}. In the context of
random-effects meta-analysis and meta-regression, this equivariance
implies that not only we get a median bias-reduced estimator of
$\psi$, but we also get median bias-reduced estimates of the standard
errors for $\beta$ by calculating the square roots of the diagonal
elements of $\{i(\theta)\}^{-1}$ in (\ref{eq:metaregexpectedinfo}) at
$\psi^\dagger$. This is because $i(\theta)$ is a function of $\psi$ only, and moreover the square
roots of the diagonal elements of $\{i(\theta)\}^{-1}$ are monotone functions of $\psi$.

\subsection{Penalized likelihood-based inference}
For inference about either the components of the fixed-effect
parameters $\beta$ or the between-study heterogeneity $\psi$ we
propose the use of the median BRPL ratio. If
$\theta=(\tau^{\T},\lambda^{\T})^{\T}$ and
$\hat{\lambda}^{\dagger}_{\tau}$ is the maximizer of
$l^{\dagger}(\theta)$ for fixed $\tau$, then the same arguments as in
\citet{kosmidis2017improving} can be used to show that the logarithm of the
median BRPL ratio statistic
\begin{equation}
2\{l^{\dagger}(\hat{\tau}^{\dagger},\hat{\lambda}^{\dagger}) - l^{\dagger}(\tau,\hat{\lambda}^{\dagger}_{\tau})\}\,.
\label{eq:MPLRstatistic}
\end{equation}
has a $\chi^{2}_{\dim(\tau)}$ asymptotic distribution, as $K$ goes to infinity. Specifically, the
adjustment to the score function is additive and of order $O(1)$. As a
result, the extra terms in the asymptotic expansion of the logarithm
of the median BRPL that depend on the penalty and its derivatives
disappear as information increases, and the expansion has the same
leading term as that of the log-likelihood \citep[see, for example,][Section 9.4]{pace1997principles}.

\section{Cocoa intake and blood pressure reduction data (revisited)}
\label{section:cocoa2}
The ML estimate, the maximum mean BRPL estimate and the maximum median
BRPL estimate of the heterogeneity parameter in the meta-analysis
model in Section~\ref{section:cocoa} are $\hat{\psi} = 4.199$,
$\hat{\psi}^{\ast} = 5.546$, and $\hat{\psi}^{\dagger} = 6.897$,
respectively. The estimates of the common effect are
$\hat{\beta} = -2.799, \hat{\beta}^{\ast} = -2.811$, and
$\hat{\beta}^{\dagger} = -2.818$, with standard errors $1.000$,
$1.128$, and $1.242$, respectively. The bias-reduced estimates of
$\psi$ and, as a consequence, the corresponding estimated standard
errors for $\beta$ are larger than their ML counterparts, which is
typical in random-effects meta-analysis. The iterative process used
for computing the ML, maximum mean BRPL, and maximum median BRPL
estimates converged in 4, 5, and 11 iterations, respectively. The
computational run-time for the two-step iterative process which
computes the ML, maximum mean BRPL, and maximum median BRPL estimates
is $1.1 \times 10^{-2}$, $1.8 \times 10^{-2}$, and
$1.1 \times 10^{-2}$ seconds, respectively.

\begin{figure}
\centering
\includegraphics[scale=0.5]{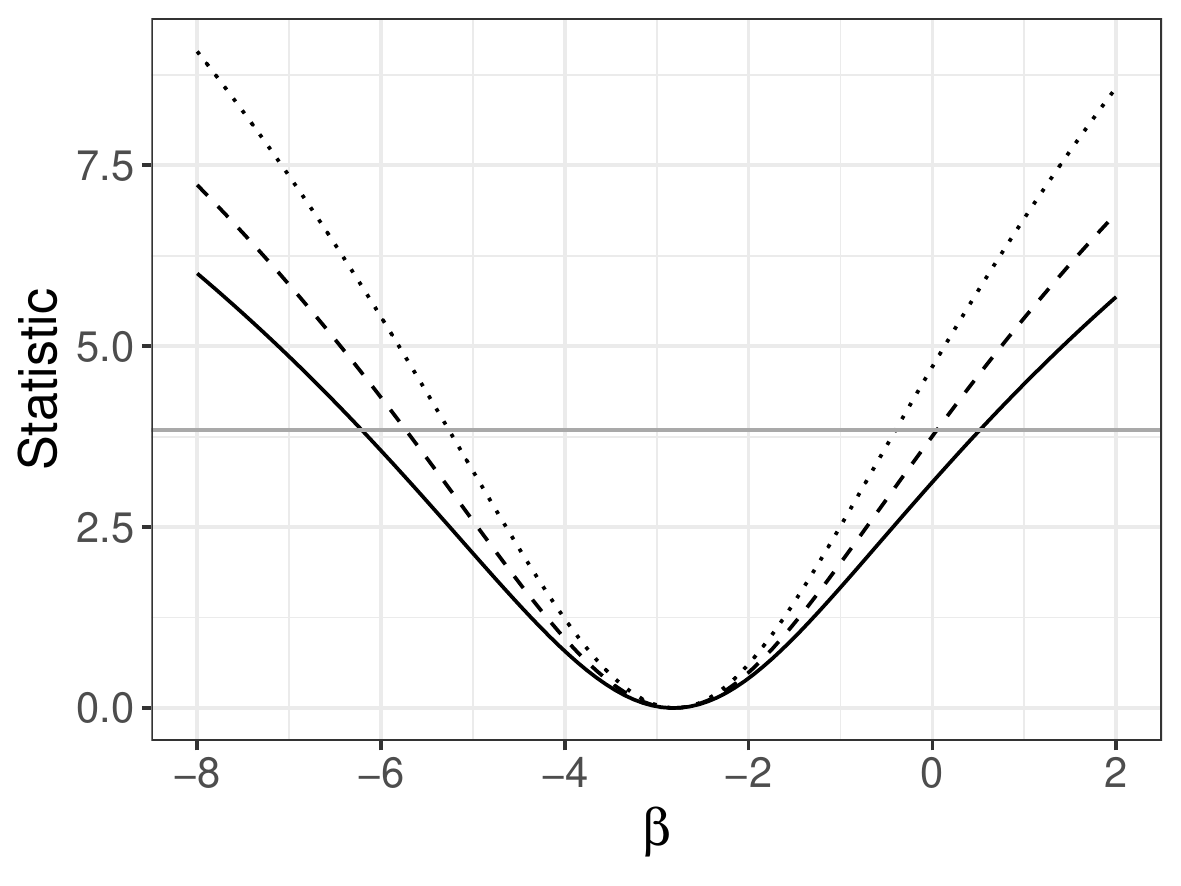}
\includegraphics[scale=0.5]{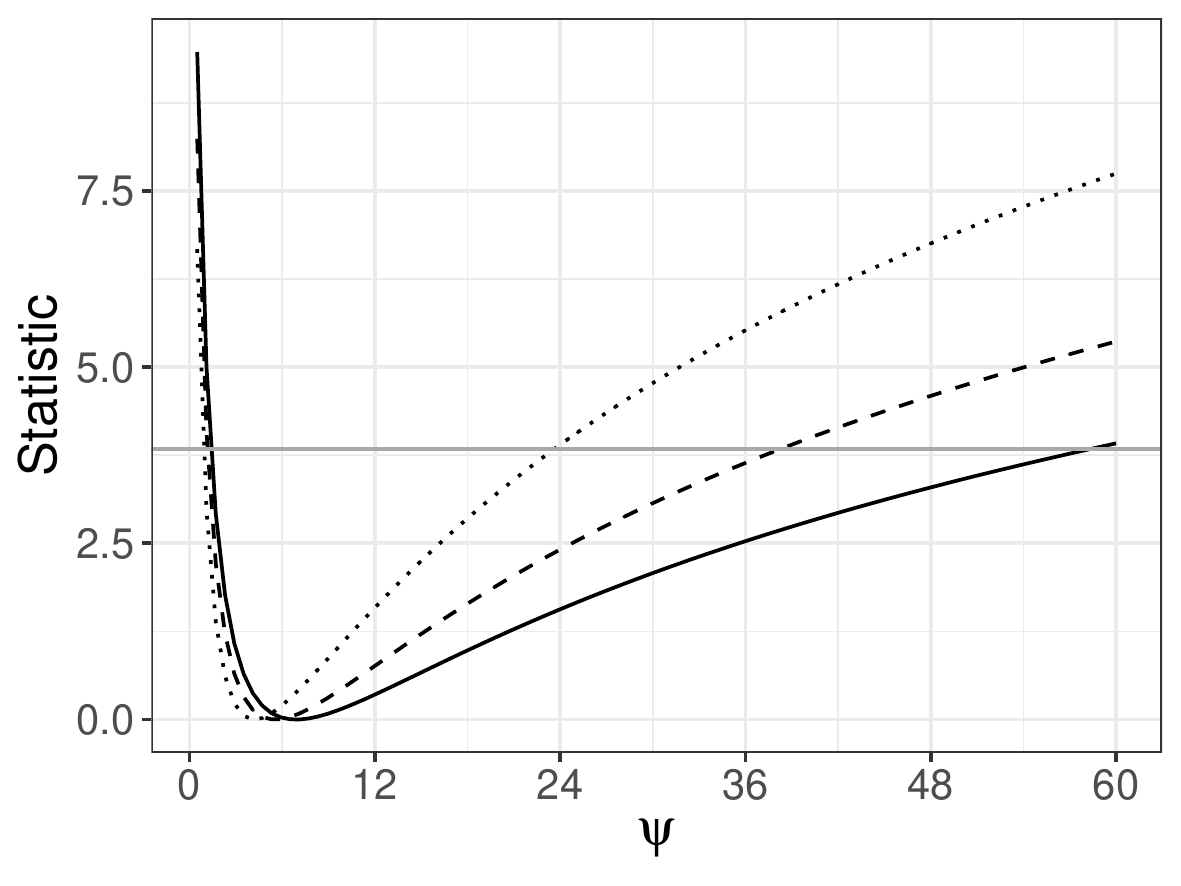}
\caption{Plot of LR (dotted), mean BRPL (dashed) and median BRPL (solid) ratio statistic in~(\ref{eq:MPLRstatistic}) when $\tau$ is $\beta$ (left) and $\psi$ (right). The horizontal line is the $95\%$ quantile of the limiting $\chi^2_1$ distribution, and its intersection with the values of the statistics results in the endpoints of the corresponding $95\%$ confidence intervals.}
\label{LRCI:cocoa}
\end{figure}

Figure~\ref{LRCI:cocoa} shows the value of LR, mean BRPL and median BRPL ratio statistic in~(\ref{eq:MPLRstatistic}) for a range of values of $\tau$, when $\tau$ is either $\beta$ or $\psi$. Here and in the following simulation studies we compare median BRPL ratio statistic with only LR and mean BRPL ratio statistics  because the mean BRPL ratio statistic is a strong competitor against other alternatives in terms of inferential performance \citep{kosmidis2017improving}. The horizontal line in Figure~\ref{LRCI:cocoa} is the $95\%$ quantile of the limiting $\chi^2_1$ distribution,
and its intersection with the values of the statistics results in the endpoints of the corresponding $95\%$ confidence intervals. For both $\beta$ and $\psi$, the confidence intervals based on the LR statistic are the narrowest and the confidence intervals based on the median
BRPL ratio statistic are the widest. Specifically, the $95\%$
confidence intervals for $\beta$ are $(-6.21,0.52)$, $(-5.73,0.05)$, and $(-5.26,-0.40)$ for the median BRPL ratio statistic, mean BRPL ratio statistic, and LR statistic, respectively. The corresponding $95\%$ confidence intervals for $\psi$ are $(1.4,58.0)$, $(1.0,38.5)$, and $(1.1,23.5)$, respectively. Contrary to the LR test, the mean BRPL and median BRPL ratio tests suggest that there is only weak evidence that cocoa consumption affects diastolic blood pressure with $p$-values of $0.053$ and $0.077$.

\begin{figure}
\centering
\includegraphics[scale=0.5]{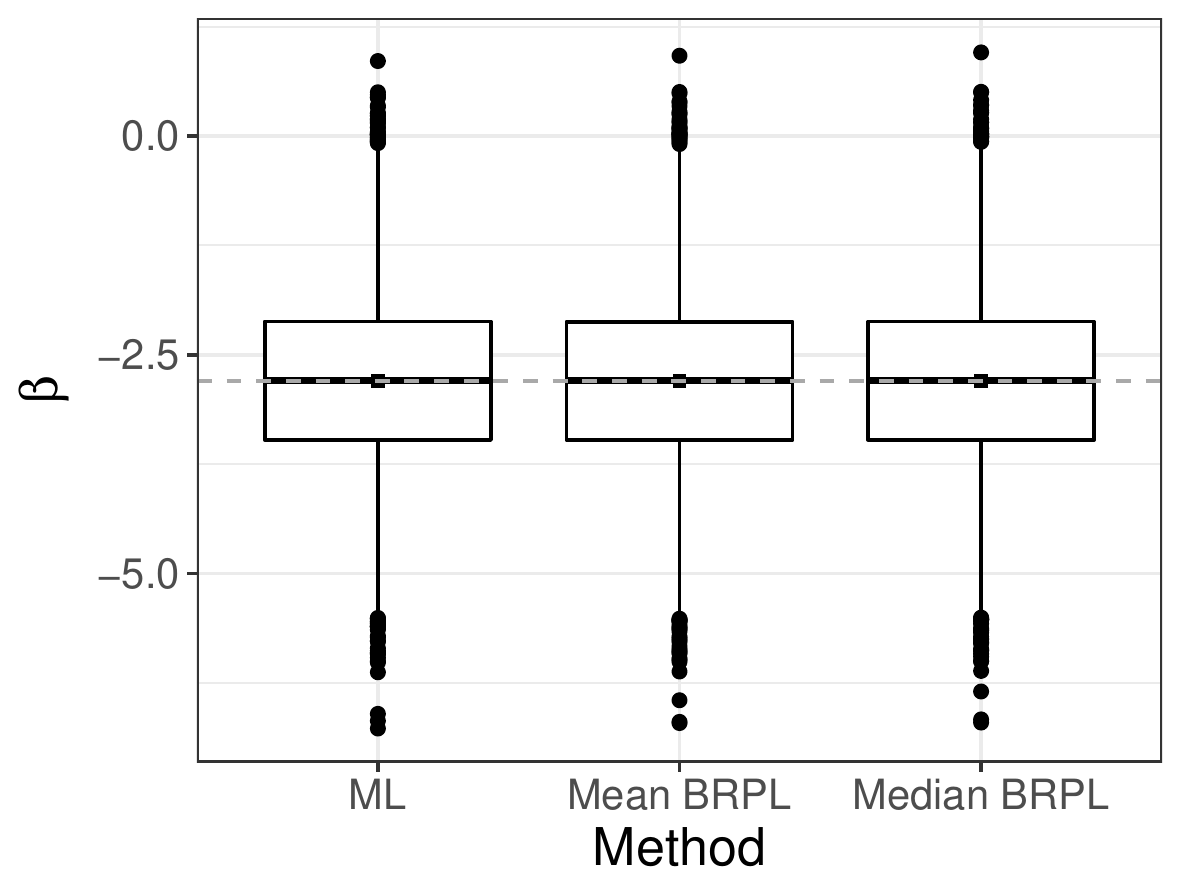}
\includegraphics[scale=0.5]{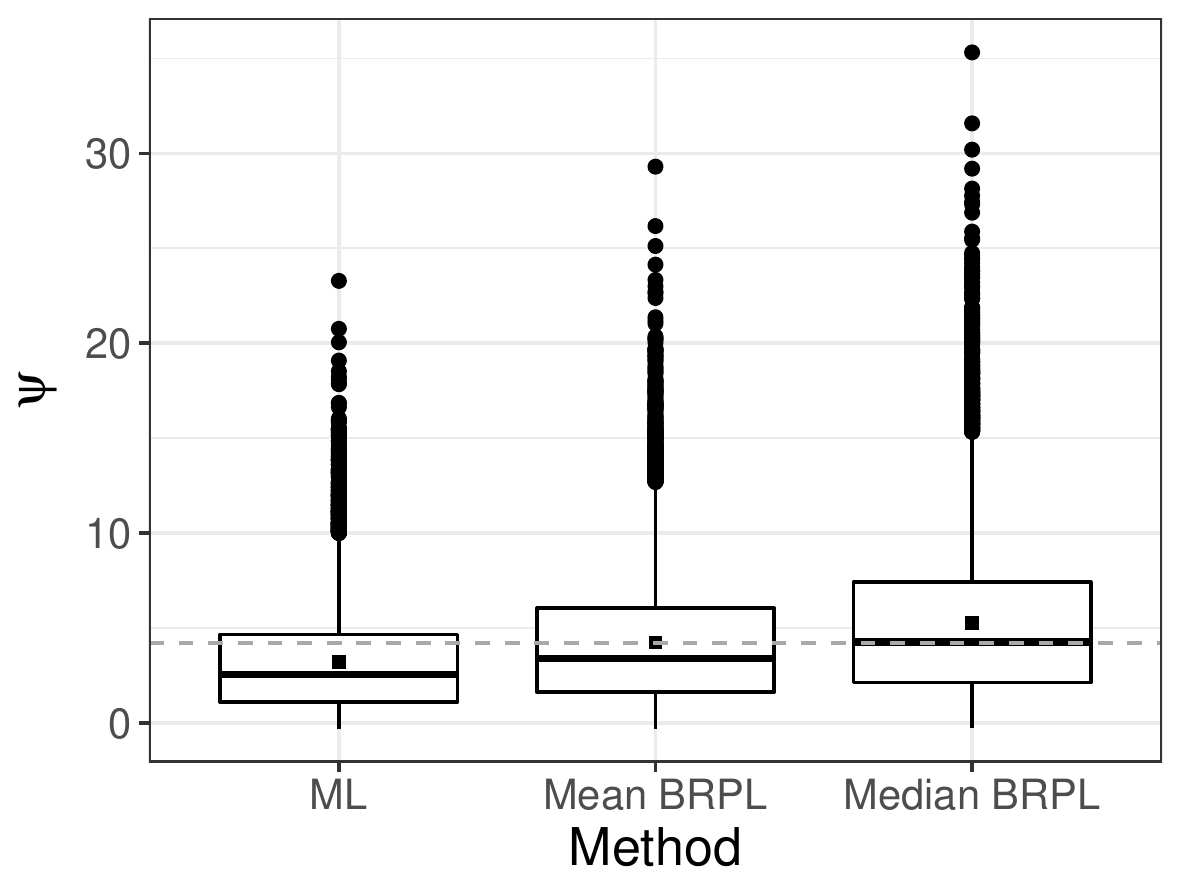}
\caption{Boxplots for the ML, the maximum mean BRPL, and the maximum median BRPL estimates of $\beta$ and $\psi$ as calculated from $10\,000$ simulated samples under the ML fit using the cocoa data \citep{bellio2016integrated, taubert2007effect}. The square point is the empirical mean of the estimates. The dashed grey horizontal line is at the parameter value used to generate the data.}
\label{boxplot:cocoa}
\end{figure}

In order to further investigate the performance of the three
approaches to estimation and inference, we performed a simulation
study where we simulated $10\,000$ independent samples from the
random-effects meta-analysis model with parameter values set to the ML
estimates reported earlier, i.e. $\beta_0 = -2.799$ and
$\psi_0 = 4.199$. Figure~\ref{boxplot:cocoa} shows boxplots of the
estimates of $\beta$ and $\psi$ calculated from each of the $10\,000$
simulated samples. The distributions of the three alternative
estimators for $\beta$ are similar. On the other hand, the ML
estimator of $\psi$ has a large negative mean bias, maximum median
BRPL tends to over-correct for that bias, while maximum mean BRPL
almost fully corrects for the bias of ML estimator. The distribution
of the median BRPL estimates has the heaviest right tail. The
simulation-based estimates of the probabilities of underestimation for
$\psi$, $P_{\psi_0}(\hat{\psi} \leq \psi_0)$,
$P_{\psi_0}(\hat{\psi}^{\ast} \leq \psi_0)$ and
$P_{\psi_0}(\hat{\psi}^{\dagger} \leq \psi_0)$ are 0.708, 0.591 and
0.493 for the ML, maximum mean BRPL, and maximum median BRPL,
respectively, illustrating how effective maximizing the median BRPL in
(\ref{eq:regpenalizedlik_median}) is in reducing the median bias of
the maximum likelihood estimator of $\psi$.

The simulated samples were also used to calculate the empirical
$p$-value distribution for the two-sided tests that each parameter is
equal to the true values based on the LR statistic, the mean BRPL
ratio statistic, and the median BRPL ratio
statistic. Table~\ref{table:cocoa_LRdistr} shows that the empirical
$p$-value distribution for the mean and median BRPL ratio statistics
are closest to uniformity, with the latter being slightly more
conservative than the former.  The coverage probability of the $95\%$
confidence intervals of $\beta$ based on the mean BRPL ratio and the
median BRPL ratio are notably closer to the nominal level than those
based on the likelihood ratio. Specifically, the coverage
probabilities for $\beta$ are $88\%$, $93\%$, and $96\%$ for LR, mean
BRPL ratio, and median BRPL ratio respectively, and the corresponding
coverage probabilities for $\psi$ are $88\%$, $94\%$, and $96\%$,
respectively.

\begin{table}
\centering
\caption{Empirical $p$-value distribution (\%) for the tests based on the LR statistic, the mean BRPL ratio statistic, and the median BRPL ratio statistic in the cocoa data \citep{bellio2016integrated, taubert2007effect} setting.}
\label{table:cocoa_LRdistr}
\begin{tabular}{lccccccccccc}
\toprule
$\alpha \times 100$ & $1.0$ & $2.5$ & $5.0$ & $10.0$ & $25.0$ & $50.0$ & $75.0$ & $90.0$ & $95.0$ & $97.5$ & $99.0$ \\
\midrule
LR & 5.7 & 8.4 & 11.8 & 18.2 & 34.6 & 57.8 & 79.2 & 91.8 & 96.1 & 98.0 & 99.3 \\
Mean BRPL ratio & 1.6 & 3.7 & 6.7 & 12.2 & 28.4 & 52.8 & 76.6 & 90.9 & 95.5 & 97.9 & 99.1 \\
Median BRPL ratio & 0.6 & 1.8 & 4.1 & 8.6 & 23.1 & 48.5 & 74.2 & 90.0 & 95.0 & 97.5 & 99.1 \\
\bottomrule
\end{tabular}
\end{table}

\section{Simulation study}
\label{section:simulation_MA}
More extensive simulations under the random-effects meta-analysis
model (\ref{eq:metamodel}) are performed here using the design in \citet{brockwell2001comparison}. Specifically, the
data $y_i\,$, $i \in \{1,\ldots,K\}$, are simulated from model
(\ref{eq:metamodel}) with true fixed-effect parameter $\beta=0.5$. The
within-study variances $\hat{\sigma}^2_i$ are independently generated
from a $\chi^{2}_{1}$ distribution and are multiplied by $0.25$ before
restricted to the interval $(0.009, 0.6)$.  Eleven values of the
between-study variance $\psi$ ranging from $0$ to $0.1$ are chosen,
and the number of studies $K$ ranges from $5$ to $200$. For each
combination of $\psi$ and $K$ considered, we simulated $10\,000$ data
sets initializing the random number generator at a common state. The
within-study variances where generated only once and kept fixed while
generating the samples.

\citet{zeng2015random} compared the performance of their proposed
double resampling method with the \citet{dersimonian1986meta} method,
the profile likelihood method in \citet{hardy1996likelihood}, and the
resampling method in \citet{jackson2009re} and showed that the double
resampling method improves the accuracy of statistical
inference. Based on these results \citet{kosmidis2017improving}
compared the performance of their mean BRPL approach with the double
resampling method and illustrated that the former results in
confidence intervals with coverage probabilities closer to the nominal
level that the alternative methods.

We take advantage of the results reported in \citet{zeng2015random}
and \citet{kosmidis2017improving} and evaluate the performance of
estimation and inference based only on the median BRPL with that based
on the likelihood and the mean BRPL. The estimators of the fixed and
random-effect parameters obtained from the three methods are
calculated using variants of the two-step algorithm described in
Section~\ref{section:algorithm}. In the second step of the algorithm
the candidate values for the ML, and maximum mean and median BRPL
estimators of the between-study variance $\psi$ are calculated by
searching for the root of the partial derivatives of $l(\theta)$,
$l^{\ast}(\theta)$, and $l^{\dagger}(\theta)$ with respect to $\psi$,
in the interval $(0,3)$.

\begin{figure}
\centering
\includegraphics[scale=0.5]{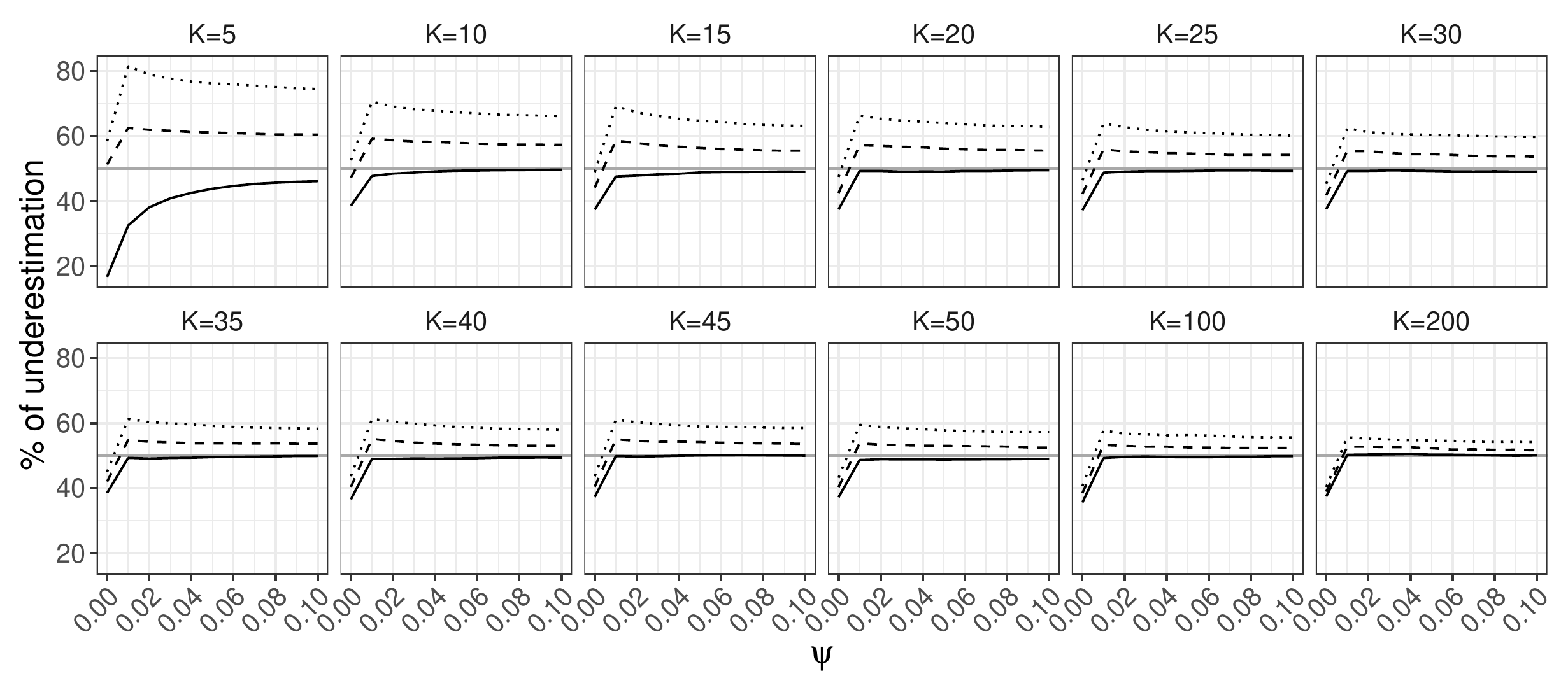}
\caption{Empirical percentage of underestimation for $\psi$ for
  random-effects meta-analysis. The percentage of underestimation is
  calculated for $K \in \{5,10,15,20,25,30,35,40,45,50,100,200\}$ and
  for increasing values of $\psi$ in the interval $[0,0.1]$. The
  curves correspond to the maximum median BRPL (solid), maximum mean
  BRPL (dashed), and ML (dotted) estimators. The grey horizontal line
  is at the target of 50\% underestimation.}
\label{fig:PU_metaAnalysis}
\end{figure}

\begin{figure}
\centering
\includegraphics[scale=0.5]{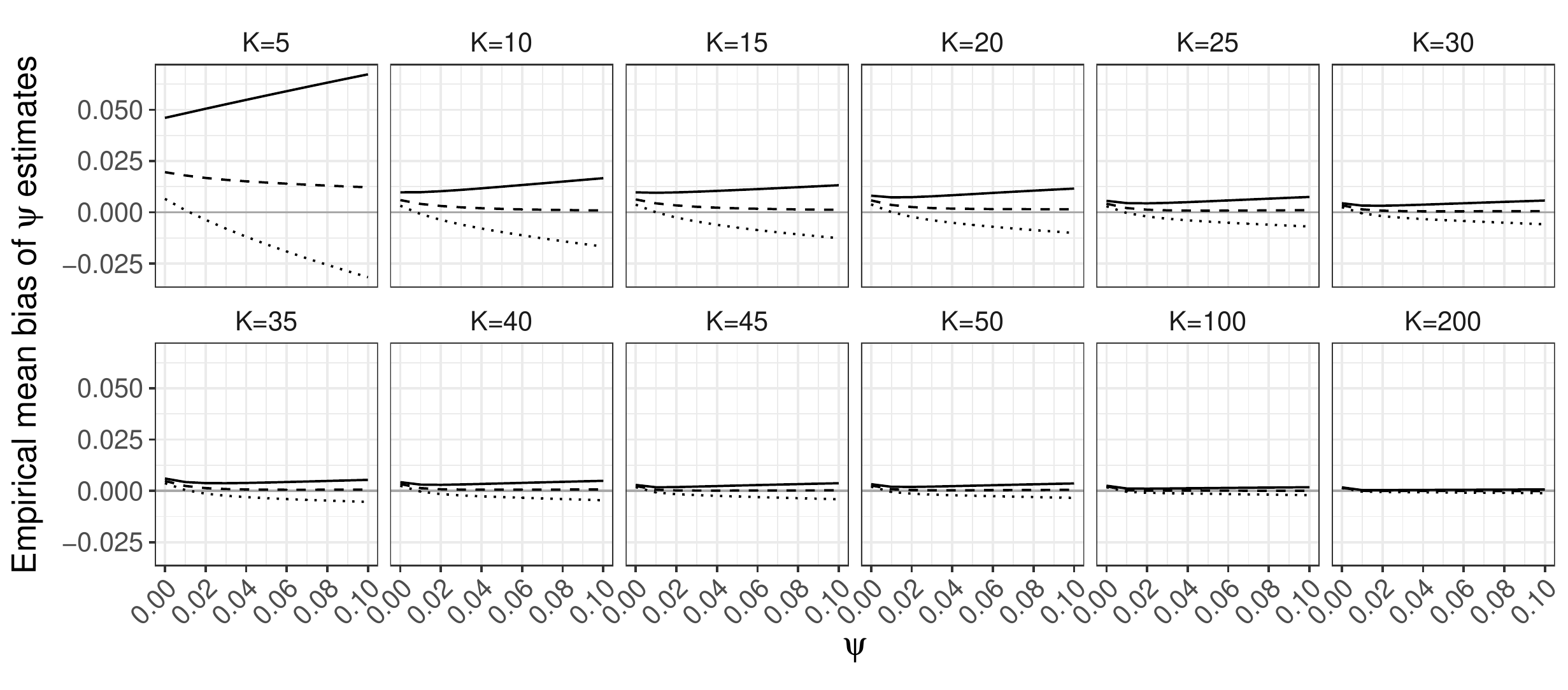}
\caption{Empirical mean bias of $\psi$ estimates for random-effects
  meta-analysis. The mean bias is calculated for
  $K \in \{5,10,15,20,25,30,35,40,45,50,100,200\}$ and for increasing
  values of $\psi$ in the interval $[0,0.1]$. The curves correspond to
  the maximum median BRPL (solid), maximum mean BRPL (dashed), and ML
  (dotted) estimators. The grey horizontal line is at zero.}
\label{fig:BIAS_metaAnalysis}
\end{figure}

First, we compare the performance of the ML, maximum mean BRPL and
maximum median BRPL estimators in terms of percentage of
underestimation. Figure~\ref{fig:PU_metaAnalysis} shows that the
median bias-reducing adjustment is the most effective in reducing
median bias even for small values of $K$. As expected, the ML and
maximum mean BRPL estimators also approach the limit of $50\%$
underestimation as $K$ grows, with the latter being closer to $50\%$
than the former. Figure~\ref{fig:BIAS_metaAnalysis} shows that maximum
median BRPL is also effective in reducing the mean bias of the ML
estimator of $\psi$ but only for moderate to large values of $K$, while maximum
mean BRPL results in estimators with the smallest bias.

\begin{figure}
\centering
\includegraphics[scale=0.5]{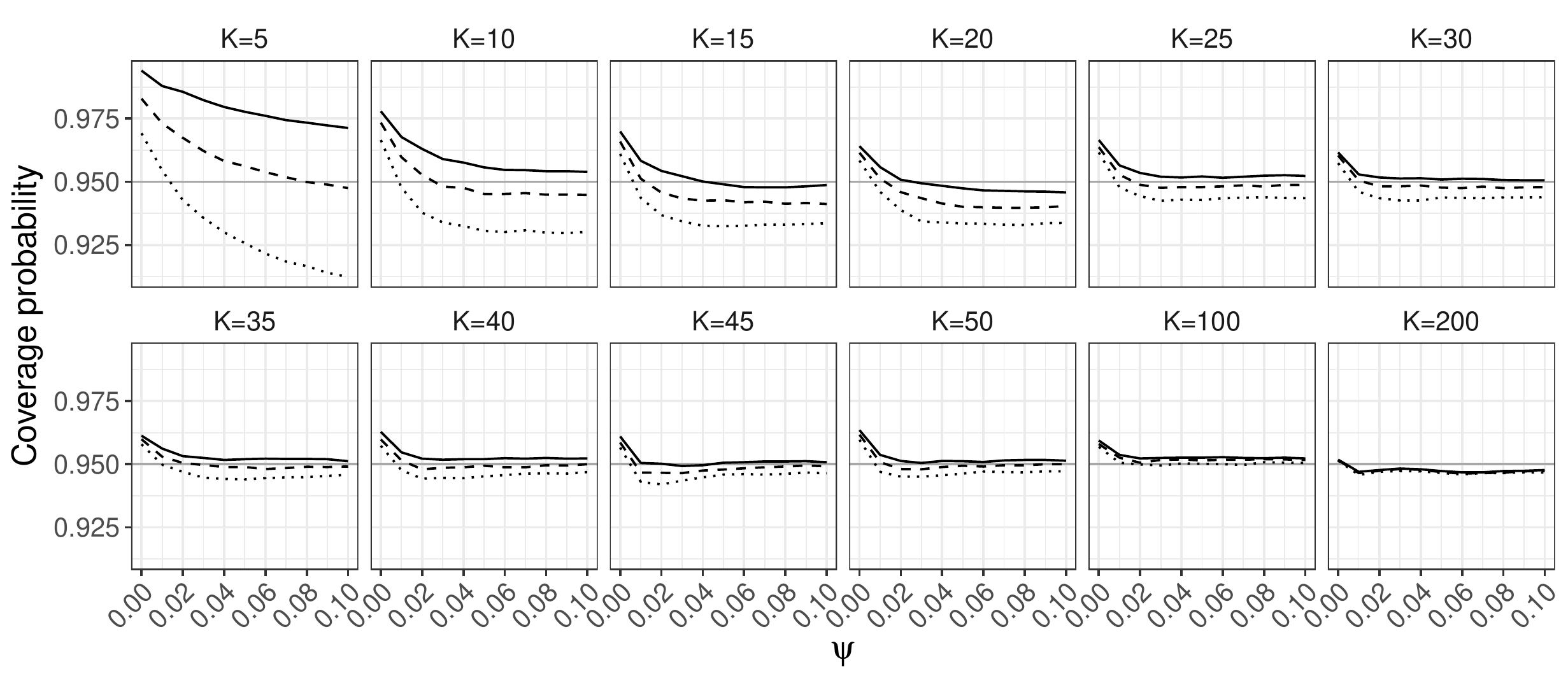}
\caption{Empirical coverage probabilities of one-sided (right)
  confidence intervals for $\beta$ for random-effects meta-analysis. The empirical
  coverage is calculated for increasing values of $\psi$ in the
  interval $[0,0.1]$ and for
  $K \in \{5,10,15,20,25,30,35,40,45,50,100,200\}$. The curves
  correspond to nominally $95\%$ internals based on the median BRPL
  ratio (solid), the mean BRPL ratio (dashed), and the LR
  (dotted). The grey horizontal line is at the $95\%$ nominal level.}
\label{fig:LRcoverage_metaAnalysis_1sided}
\end{figure}

\begin{figure}
\centering
\includegraphics[scale=0.5]{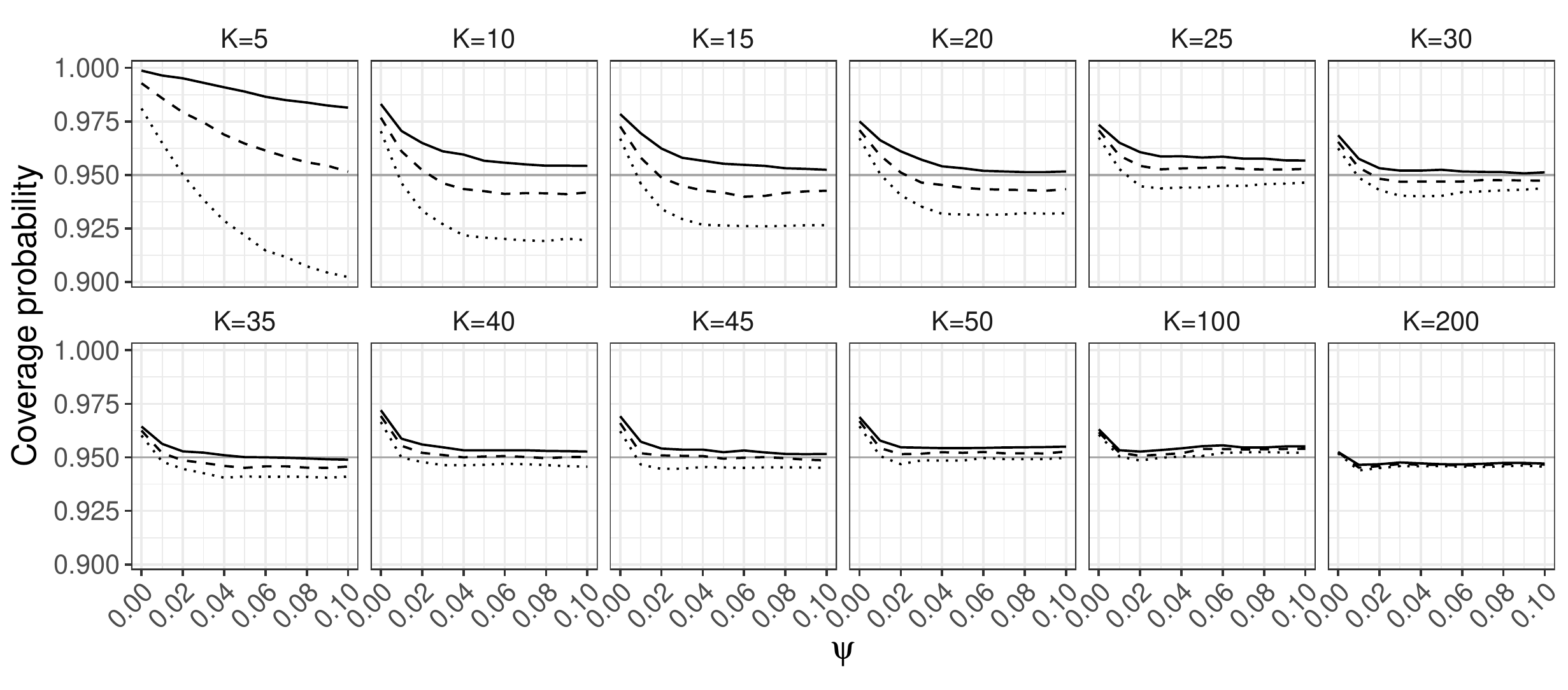}
\caption{Empirical coverage probabilities of two-sided confidence
  intervals for $\beta$ for random-effects meta-analysis. The empirical coverage
  is calculated for increasing values of $\psi$ in the interval
  $[0,0.1]$ and for $K \in
  \{5,10,15,20,25,30,35,40,45,50,100,200\}$. The curves correspond to
  nominally $95\%$ confidence intervals based on the median BRPL ratio
  (solid), the mean BRPL ratio (dashed), and the LR (dotted). The grey
  horizontal line is at the $95\%$ nominal level.}
\label{fig:LRcoverage_metaAnalysis_2sided}
\end{figure}

\begin{figure}
\centering
\includegraphics[scale=0.5]{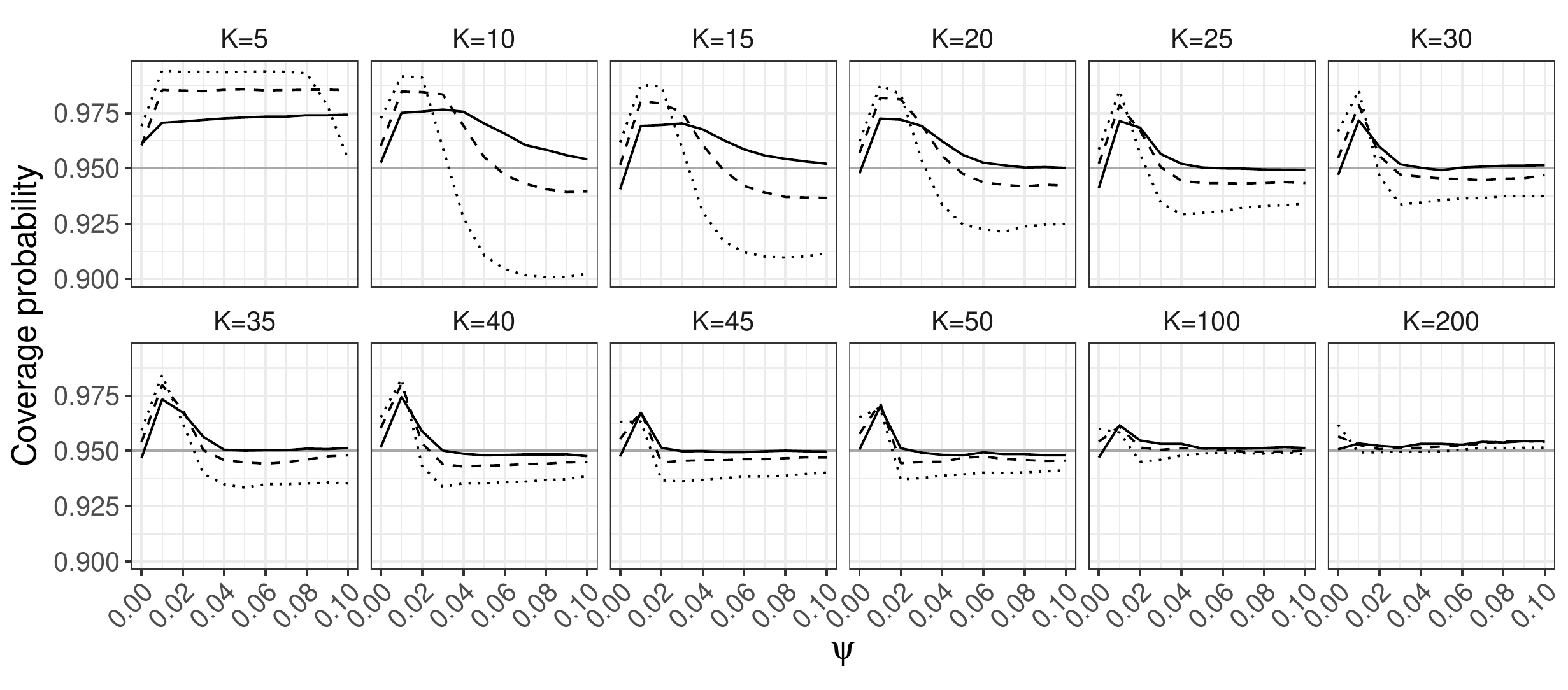}
\caption{Empirical coverage probabilities of two-sided confidence
  intervals for $\psi$ for random-effects meta-analysis. The empirical coverage
  is calculated with $\beta = 0.5$ and for increasing values of $\psi$ in the interval
  $[0,0.1]$ and for $K \in
  \{5,10,15,20,25,30,35,40,45,50,100,200\}$. The curves correspond to
  nominally $95\%$ confidence intervals based on the median BRPL ratio
  (solid), the mean BRPL ratio (dashed), and the LR (dotted). The grey
  horizontal line is at the $95\%$ nominal level.}
\label{fig:PSIcoverage_metaAnalysis}
\end{figure}

Figures~\ref{fig:LRcoverage_metaAnalysis_1sided} and
\ref{fig:LRcoverage_metaAnalysis_2sided} show the estimated coverage probability for the one-sided and two-sided confidence intervals for $\beta$ based on the LR, mean BRPL ratio and median BRPL ratio statistics at the $95\%$ nominal level. Figure~\ref{fig:PSIcoverage_metaAnalysis} shows the estimated coverage probability for the two-sided confidence intervals for $\psi$ based on the LR, mean BRPL ratio and median BRPL ratio statistics at the $95\%$ nominal level. For small values of $\psi$ or small and moderate number of studies $K$ the empirical coverage of the intervals is larger than the nominal $95\%$ level. In general, the confidence intervals based on mean and median BRPL ratio have empirical coverage that is closer to the nominal level with the latter having generally better coverage. The differences between the three methods diminish as the number of studies $K$ increases.

\begin{figure}
\centering
\includegraphics[scale=0.5]{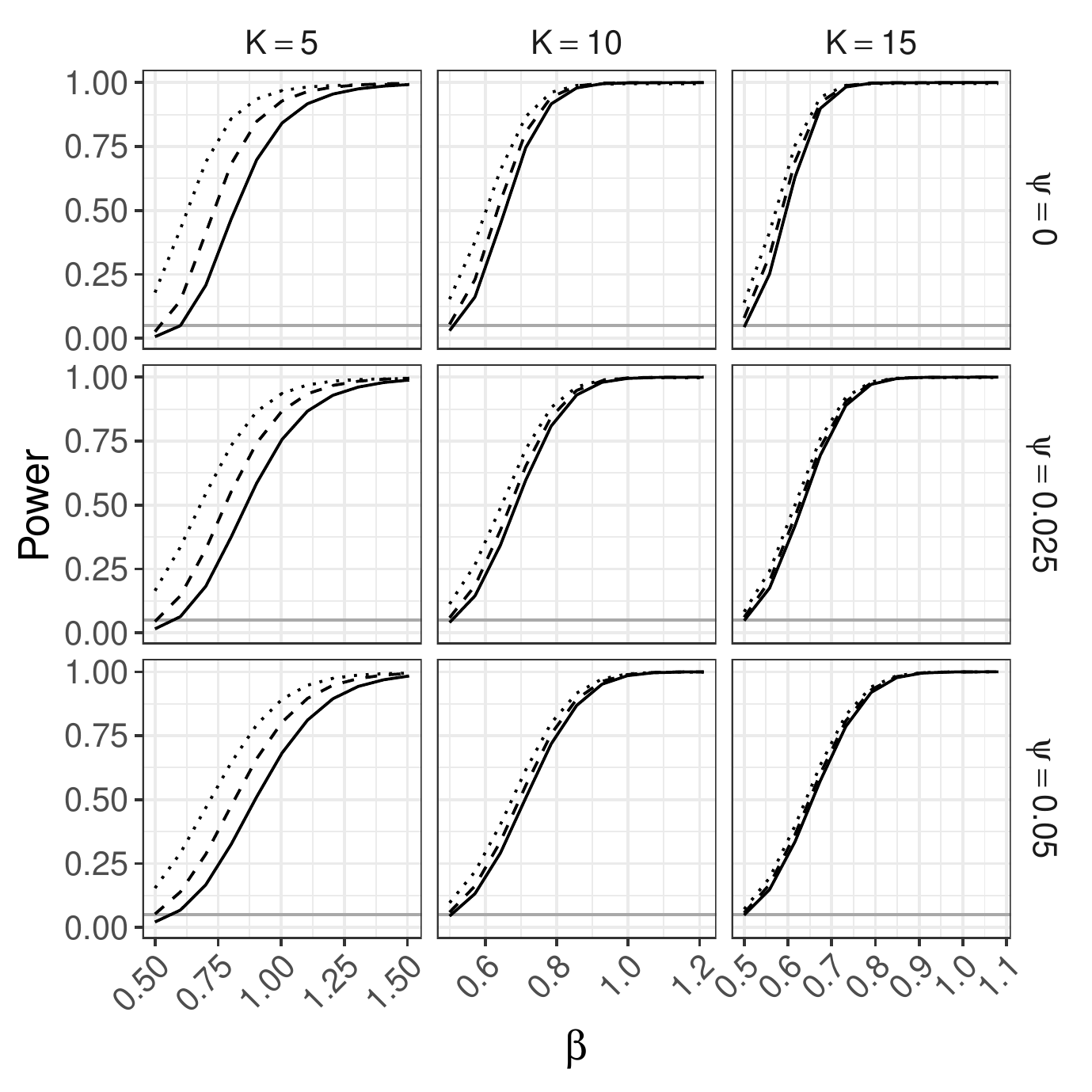}
\caption{Empirical power of the likelihood-based tests of asymptotic
  level $0.05$ for random-effects meta-analysis for testing
  $\beta=0.5$. The empirical power is calculated for increasing values
  of $\beta$, for $K \in \{5,10,15\}$ and $\psi \in
  \{0,0.025,0.05\}$. The curves correspond to median BRPL ratio
  (solid), mean BRPL ratio (dashed), and LR (dotted) tests. The grey
  horizontal line is at the $5\%$ nominal size.}
\label{fig:LRpower_metaAnalysis_allK_asympt}
\end{figure}

\begin{figure}
\centering
\includegraphics[scale=0.5]{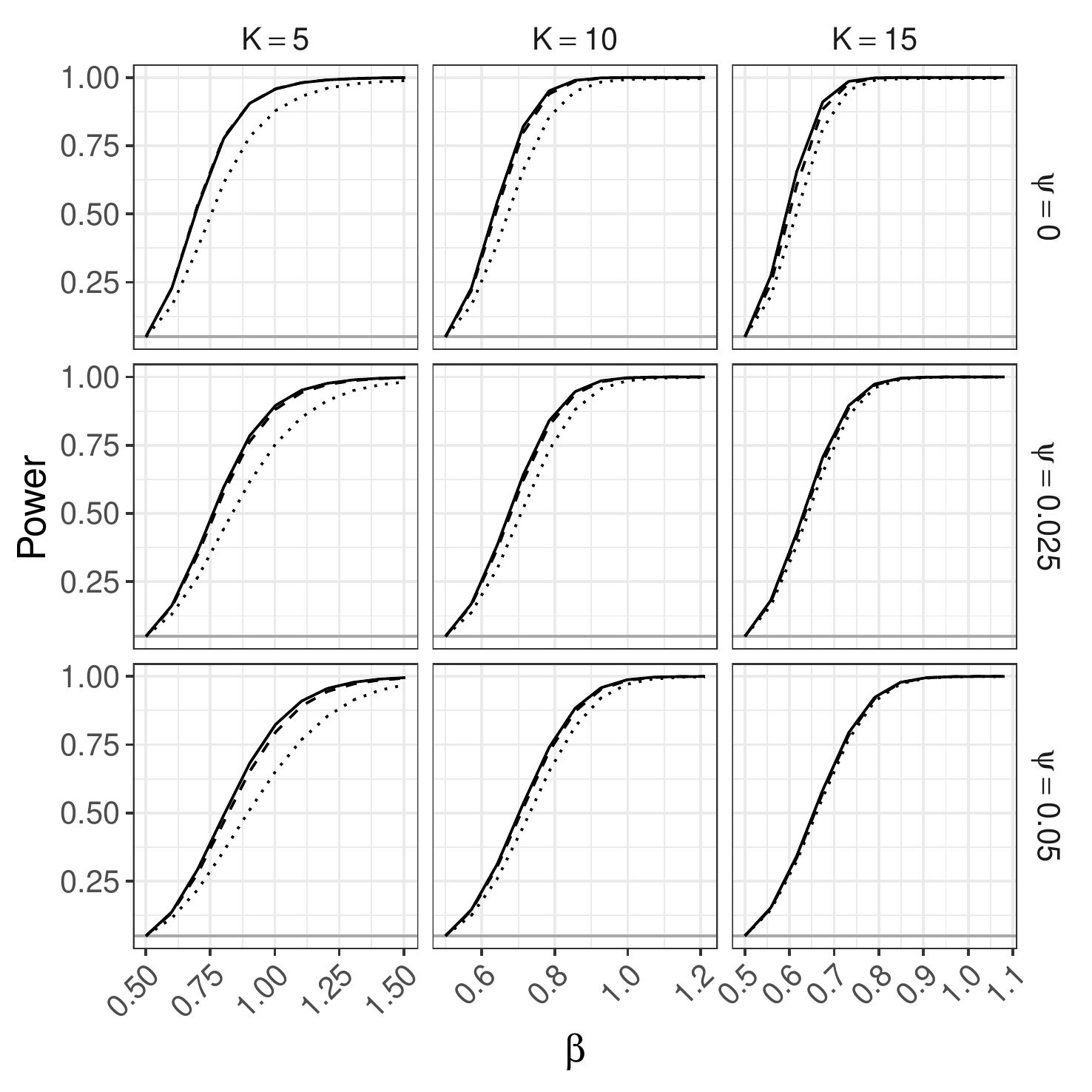}
\caption{Empirical power of the likelihood-based tests of exact level
  $0.05$ for random-effects meta-analysis for testing $\beta=0.5$. The
  empirical power is calculated for increasing values of $\beta$, for
  $K \in \{5,10,15\}$ and $\psi \in \{0,0.025,0.05\}$. The curves
  correspond to median BRPL ratio (solid), mean BRPL ratio (dashed),
  and LR (dotted) tests. The grey horizontal line is at the $5\%$
  nominal size.}
\label{fig:LRpower_metaAnalysis_allK_exact}
\end{figure}

Figures~\ref{fig:LRpower_metaAnalysis_allK_asympt} and
\ref{fig:LRpower_metaAnalysis_allK_exact} give the power of the LR,
the mean BRPL ratio, and the median BRPL ratio tests for testing the
null hypothesis $\beta = 0.5$ against various
alternatives. Specifically, we simulated $10~000$ data sets under the
alternative hypothesis that parameter $\beta$ is equal to
$b = 0.5 + \delta K^{-1/2}$, where $\delta$ ranges from $0$ to
$2.25$. In Figure~\ref{fig:LRpower_metaAnalysis_allK_asympt} the power
is calculated using critical values of the the asymptotic null
$\chi^2_{1}$ distribution of the statistics. In
Figure~\ref{fig:LRpower_metaAnalysis_allK_exact} the power is
calculated using critical values based on the exact null distribution
of each statistic, obtained by simulation under the null
hypothesis. In this way, the three tests are calibrated to have size
$5\%$.

Figure~\ref{fig:LRpower_metaAnalysis_allK_asympt} shows that the three
tests have monotone power and for small values of $K$ the LR test
yields the largest power. This is because the LR test is oversized,
while the mean and median BRPL ratio tests are slightly more
conservative and this conservativeness comes at the cost of lower
power. As the number of studies $K$ increases the three tests approach
the nominal size and provide similar power. The use of the exact
critical values in Figure~\ref{fig:LRpower_metaAnalysis_allK_exact}
allows us to compare the performance of the tests without letting the
oversizing or the conservativeness of a test skew the power
results. Figure~\ref{fig:LRpower_metaAnalysis_allK_exact} shows that
the power of the median BRPL ratio test is almost identical to that of
the mean BRPL ratio test, and both tests have larger power than the LR
test. Again, inference based on either of the two penalized
likelihoods becomes indistinguishable from classical likelihood
inference as the number of studies increases.

Across all $\psi$ and $K$ values considered, the average
  number of iterations taken per fit for the two-step iterative
  process to converge is 6.20, 5.75 and 5.86 iterations for ML,
  maximum mean BRPL, and maximum median BRPL, respectively. The
  average computational run-times for ML, maximum mean BRPL, and
  maximum median BRPL are 0.005, 0.021, and 0.017 seconds,
  respectively. Figures~1 and 2 in the Supplementary material show the
  average number of iterations and the average computational run-time
  taken per fit for the two-step iterative process to converge for
  each value of $K$ and $\psi$ used in the simulation study. The
  results show that in all cases estimation is achieved rapidly and
  after a small number of iterations for all three methods, with only
  negligible overhead with the two bias reducing methods.

  \section{Meat consumption data}
\label{section:meat}
\begin{figure}
\centering
\includegraphics[scale=0.5]{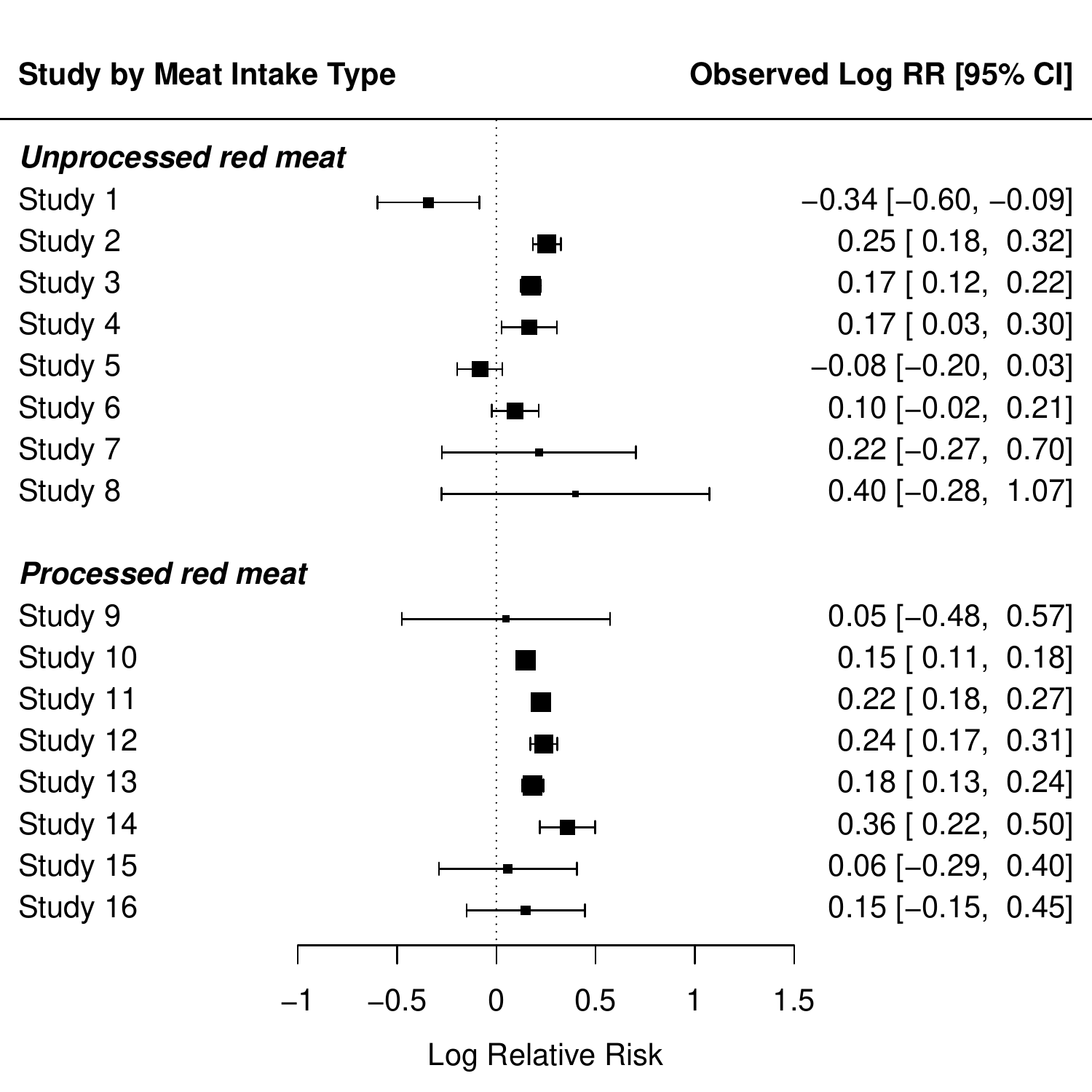}
\caption{The meat consumption data \citep{larsson2014red}. Outcomes from 16 studies are reported in terms of the logarithm of the relative risk (Log RR) of all-cause mortality for the highest versus lowest category of unprocessed red meat, and processed meat
consumption. Squares represent the mean effect estimate for each study; the size of the square reflects the weight that the corresponding study exerts in the meta-analysis. Horizontal lines represent $95\%$ Wald-type confidence intervals (CI) of the effect estimate of individual studies.}
\label{forest:meat}
\end{figure}

\citet{larsson2014red} investigate the association between meat
consumption and relative risk of all-cause mortality. The data
consists of $16$ prospective studies, eight of which are about
unprocessed red meat consumption and eight about processed meat
consumption. Figure~\ref{forest:meat} displays the information
provided by each study in the meta-analysis. The results from the
studies point towards the conclusion that high consumption of red
meat, in particular processed red meat, is associated with higher
all-cause mortality.

We consider the random-effects meta-regression model, assuming that
$Y_i$ has a $N(\beta_0 + \beta_1x_i,\hat{\sigma}^2_{i} + \psi)$, where
$Y_i$ is the random variable representing the logarithm of the
relative risk reported in the $i$th study, and $x_i$ takes value $1$
if the consumption in the $i$th study is about processed red meat and
$0$ if it is about unprocessed meat $(i = 1,\ldots,16)$.
Table~\ref{table:meatestimates} gives the ML estimates, the mean BRPL
estimates, and the median BRPL estimates of $\beta_0$, $\beta_1$ and
$\psi$, along with the corresponding standard errors for $\beta_0$ and
$\beta_1$. The median BRPL estimate of $\psi$ and the standard errors
of the fixed-effect parameters are the largest. The iterative process
for computing the ML, maximum mean BRPL, and maximum median BRPL
estimates converged in eight, nine, and twelve iterations, in
$1.2 \times 10^{-2}$, $2.4 \times 10^{-2}$, and $1.5 \times 10^{-2}$
seconds, respectively.

The LR test indicates some evidence for a higher risk associated to
the consumption of red processed meat with a $p$-value of $0.047$. On
the other hand, the mean and median BRPL ratio tests suggest that
there is weaker evidence for higher risk, with $p$-values of $0.066$
and $0.074$, respectively. Skovgaard's test also gives weak evidence
for higher risk with $p$-value 0.073.

\begin{table}
\centering
\caption{ML, maximum mean BRPL, and maximum median BRPL estimates of the model
  parameters for the meat consumption data \citep{larsson2014red}.
  Standard errors are reported in parentheses. The $95\%$ confidence
  intervals based on the LR, mean BRPL ratio and median BRPL ratio are
  reported in squared brackets.}
\label{table:meatestimates}
\begin{tabular}{cccc}
\toprule
Method & $\beta_0$ & $\beta_1$ & $\psi$\\
\midrule
ML   & 0.099 (0.044) & 0.106 (0.061) & 0.009  \\
 & [-0.004,0.189] & [-0.022,0.244] & [0.003,0.030]  \\
Maximum mean BRPL  & 0.095 (0.050) & 0.110 (0.069) & 0.012  \\
 & [-0.020,0.199] & [-0.040,0.264] & [0.003,0.042]  \\
Maximum median BRPL & 0.093 (0.052) & 0.111 (0.072) & 0.013 \\
 & [-0.027,0.203] & [-0.048,0.271] & [0.004,0.048]  \\
\bottomrule
\end{tabular}
\end{table}

\begin{figure}
\centering
\includegraphics[scale=0.5]{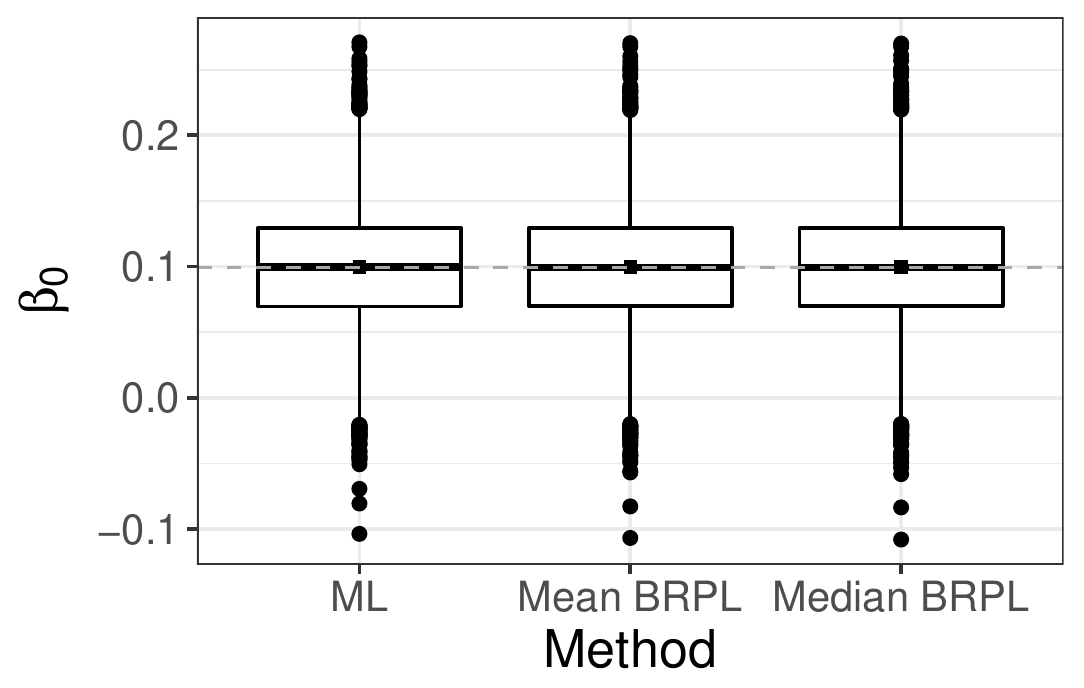}
\includegraphics[scale=0.5]{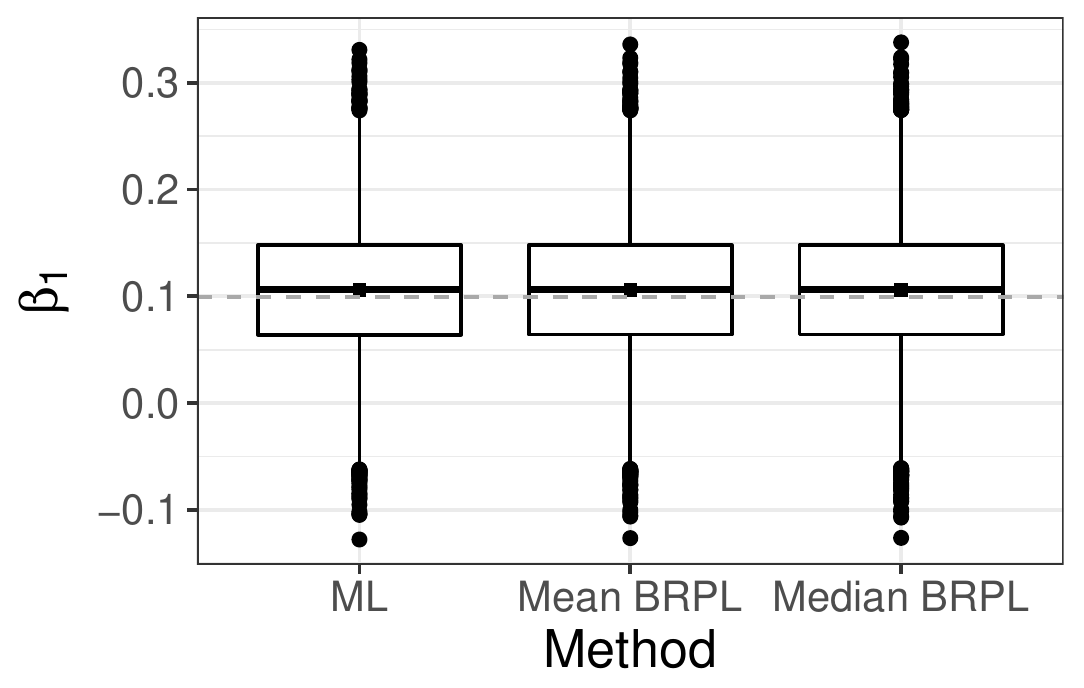}
\includegraphics[scale=0.5]{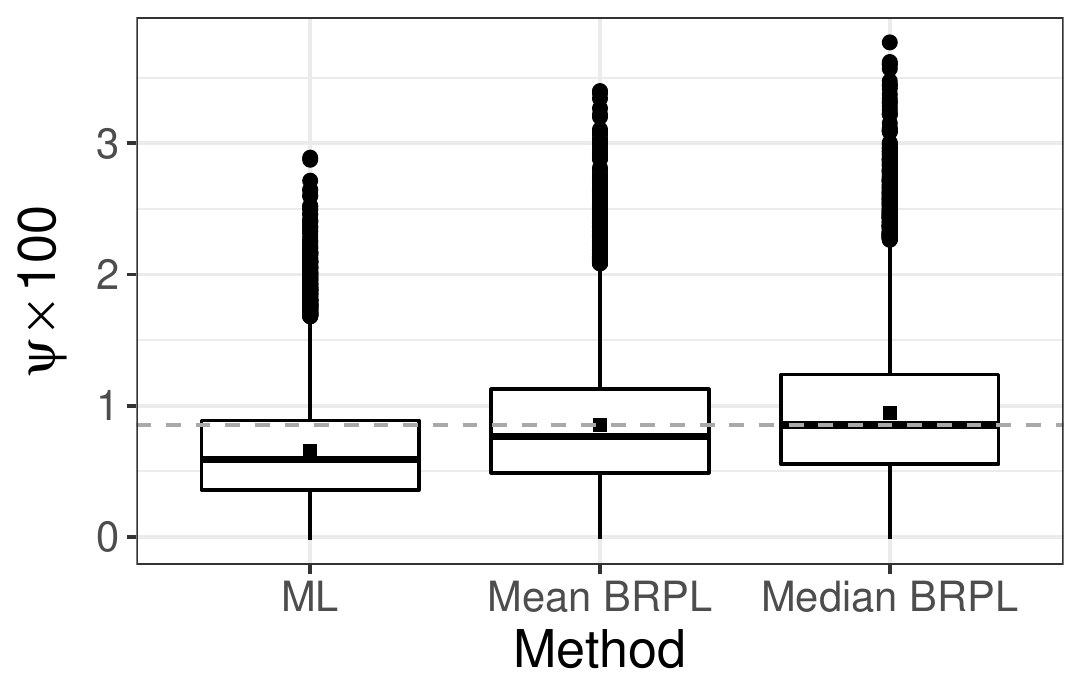}
\caption{Boxplots for the ML, maximum mean BRPL, and maximum median BRPL estimates of $\beta_0$, $\beta_1$, and $\psi$ as calculated from $10\,000$ simulated samples under the ML fit using the meat consumption data \citep{larsson2014red}. The square point is the mean of the estimates obtained from each method. The dashed grey horizontal line is at the parameter value used to generate the data.}
\label{boxplot:meat}
\end{figure}

Similar to Section~\ref{section:cocoa2}, we performed a simulation
study in order to further investigate the performance of the three
methods in a meta-regression context. We simulated $10\,000$
independent samples from the meta-regression model at the ML estimates
reported in Table~\ref{table:meatestimates}. Figure~\ref{boxplot:meat}
shows boxplots of the estimates of $\beta_0$, $\beta_1$, and
$\psi$. Maximum likelihood underestimates the parameter $\psi$, while
mean BRPL and median BRPL almost fully compensate for the negative
bias of ML estimates, with the latter having a slightly heavier right
tail. The percentages of underestimation are $72.6\%$, $56.6\%$, and
$49.9\%$ for the ML, maximum mean BRPL, and maximum median BRPL
estimators, respectively.

The simulated samples were also used to calculate the empirical
$p$-value distribution for the tests based on the likelihood, mean
BRPL and median BRPL ratio statistics. Table~\ref{table:meat_LRdistr}
shows that the empirical $p$-value distribution for the median BRPL
ratio statistic is the one closest to uniformity.

\begin{table}
\centering
\caption{Empirical $p$-value distribution (\%) for the tests based on the LR statistic, the mean BRPL ratio statistic, and the median BRPL ratio ratio statistic using the meat consumption data \citep{larsson2014red}.}
\label{table:meat_LRdistr}
\begin{tabular}{lccccccccccc}
\toprule
$\alpha \times 100$ & $1.0$ & $2.5$ & $5.0$ & $10.0$ & $25.0$ & $50.0$ & $75.0$ & $90.0$ & $95.0$ & $97.5$ & $99.0$ \\
\midrule
LR  & 2.2 & 4.5 & 7.7 & 13.1 & 28.0 & 50.0 & 71.7 & 86.6 & 92.1 & 95.3 & 97.7 \\
Mean BRPL ratio & 1.3 & 3.0 & 5.6 & 11.1 & 25.9 & 49.8 & 73.8 & 89.0 & 94.2 & 96.9 & 98.6 \\
Median BRPL ratio & 1.0 & 2.5 & 4.9 & 9.9 & 25.1 & 49.7 & 74.7 & 89.8 & 94.8 & 97.5 & 98.9 \\
\bottomrule
\end{tabular}
\end{table}

\section{Concluding remarks}
\label{section:conclusions}
In this paper we derive the adjusted score equations for the median
bias reduction of the ML estimator for random-effects meta-analysis
and meta-regression models and describe the associated inferential
procedures.

We show that the solution of the median bias-reducing adjusted score
equations is equivalent to maximizing a penalized log-likelihood. The
logarithm of that penalized likelihood differs from the logarithm of
the mean BRPL in \citet{kosmidis2017improving} by a simple additive
term. The computation of the maximum median BRPL estimators can be
performed through a two-step iteration that involves a weighted least
squares update and the solution of a nonlinear equation with respect
to a scalar parameter, and which converges rapidly, as
  illustrated by the computational times and number of iterations
  reported in the paper. The reported times and number of iterations
  were computed using a workstation with 24 cores at 2.90GHz and 80GB
  memory running under the CentOS 7 operating system, using one core
  per data set.

Using various settings we were able to retrieve enough information on
the performance of the maximum median BRPL estimators. All our
simulation studies illustrate that use of the median BRPL succeeds in
achieving median centering in estimation, and results in confidence
intervals with good coverage properties. Furthermore, while tests
based on the LR suffer from size distortions, the median BRPL ratio
statistic results in tests with size and power properties, sometimes
better to those of the mean BRPL ratio statistic in
\citet{kosmidis2017improving}.

The main advantage of the maximum median BRPL estimators from the
maximum mean BRPL ones is their equivariance under monotone
component-wise parameter transformations, which, in the case of
random-effects meta-regression, leads to median bias-reduced standard
errors.

As random-effect models are widely used in practice, the median BRPL
method is likely to be useful in models with more complex
random-effect structures, such as linear mixed models.

\section*{Supplementary material}
The \verb+R+ code for replication of examples is provided as online supplementary material.

\section*{Acknowledgements}
The authors thank E.C. Kenne Pagui for providing the formula for the median bias-reducing adjustment in matrix form.

\section*{Funding}

  The first author was partially supported by the Erasmus+ programme of the European Union which funded her for a 2-month traineeship at the University of Padova. The work of the second author was supported by the Alan Turing Institute under the EPSRC grant
  EP/N510129/1 (Turing award number TU/B/000082) and part of it was  completed when he was a Senior Lecturer at University College London. The third author was supported by the Italian Ministry of Education under the PRIN 2015 grant 2015EASZFS\_003, and by the
  University of Padova (PRAT 2015).

\section*{Appendix}

The observed information matrix for the random-effects meta-regression model (\ref{eq:metamodel}) is
\begin{equation*}
j(\theta) = \begin{pmatrix}
X^{\T}W(\psi)X & X^{\T}W(\psi)^2 R(\beta) \\
X^{\T}W(\psi)^2 R(\beta) & R(\beta)^{\T}W(\psi)^3R(\beta) - \frac{1}{2}\Tr[W(\psi)^2]
\end{pmatrix}.
\end{equation*}
For this model
\begin{equation*}
P_{t}(\theta) = -Q_{t}(\theta) =
\begin{pmatrix}
0_{p \times p} & X^{\T}W(\psi)^2X_{t} \\
X^{\T}W(\psi)^2X_{t} & 0
\end{pmatrix} \quad (t=1,\ldots,p)\,,
\end{equation*}
and
\begin{equation*}
P_{p+1}(\theta) =
\begin{pmatrix}
X^{\T}W(\psi)^2X & 0_{p} \\
0_{p}^{\T} & \Tr(W(\psi)^3)
\end{pmatrix} ~ ~ ~ \text{ and } ~ ~ ~ Q_{p+1}(\theta)=
\begin{pmatrix}
0_{p \times p} & 0_{p} \\
0_{p}^{\T} & -\Tr(W(\psi)^3)
\end{pmatrix}\,,
\end{equation*}
where $0_{p \times p}$ is the $p \times p$ zero matrix and $X_{t}$ is the $t$th column of $X$. The median bias-reducing adjustment for $\theta$ is obtained by plugging the above expressions into~(\ref{eq:medianBRadj}). The sum $P_{t}(\theta) + Q_{t}(\theta)$ ($t = 1,\ldots,p+1$) is also given in the Appendix of \citet{kosmidis2017improving}.

\includepdf[pages={1-8}]{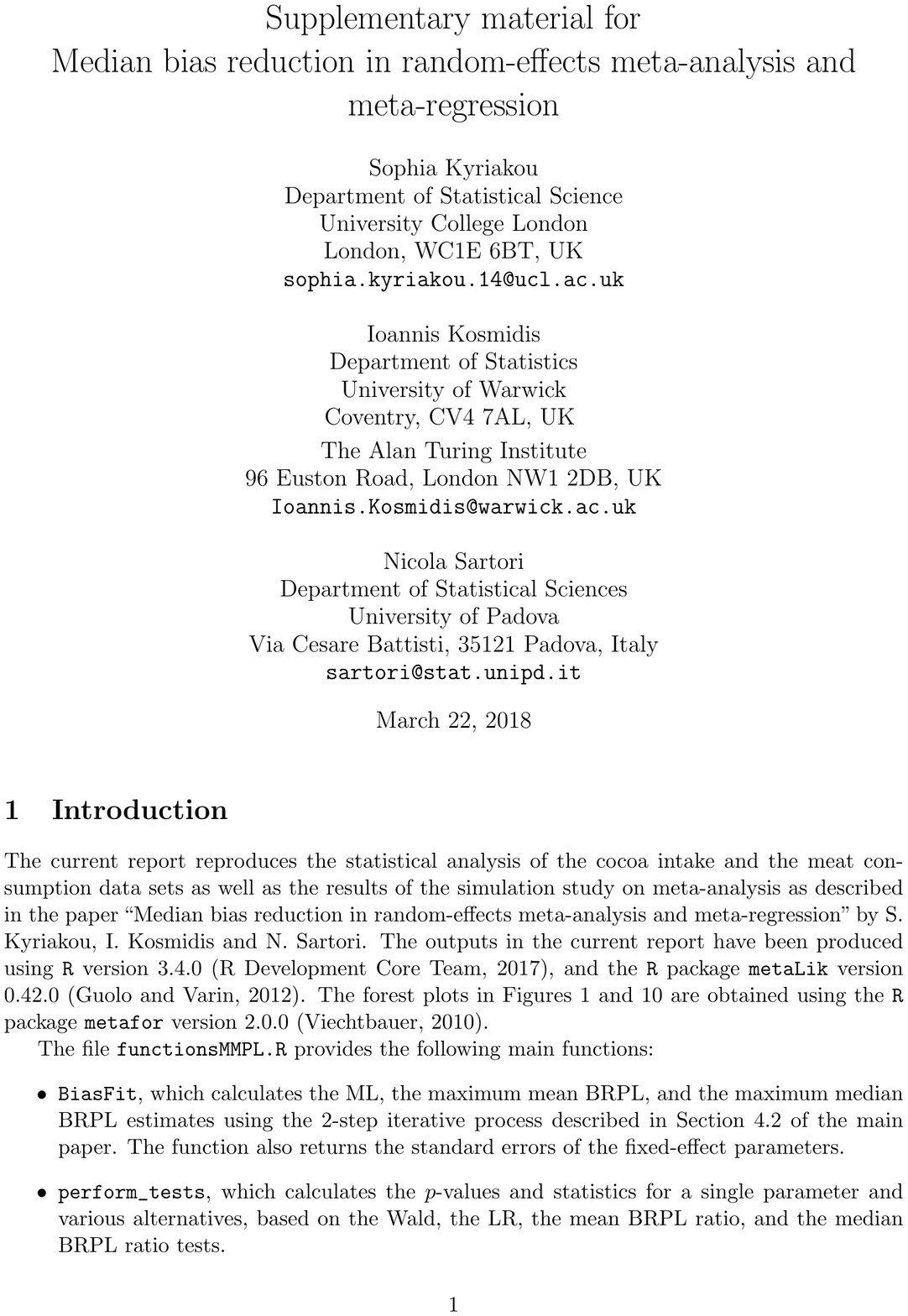}


\begin{thebibliography}{99}

\bibitem[{Bellio \& Guolo(2016)}]{bellio2016integrated}
Bellio~R and Guolo~A.
Integrated likelihood inference in small sample meta-analysis for continuous outcomes.
\textit{Scand. J. Stat.} 2016; \textbf{43}: 191--201.

\bibitem[{Brockwell \& Gordon(2001)}]{brockwell2001comparison}
Brockwell~SE and Gordon~IR.
A comparison of statistical methods for meta-analysis.
\textit{Stat. Med.} 2001; \textbf{20}: 825--840.

\bibitem[{DerSimonian \& Laird(1986)}]{dersimonian1986meta}
DerSimonian~R and Laird~N.
Meta-analysis in clinical trials.
\textit{Control. Clin. Trials} 1986; \textbf{7}: 177--188.

\bibitem[{Firth(1993)}]{firth1993bias}
Firth~D.
Bias reduction of maximum likelihood estimates.
\textit{Biometrika} 1993; \textbf{80}: 27--38.

\bibitem[{Guolo \& Varin(2015)}]{guolo2015random}
Guolo~A and Varin~C.
Random-effects meta-analysis: The number of studies matters.
\textit{Stat. Methods Med. Res.} 2017; \textbf{26}: 1500--1518.

\bibitem[{Hardy \& Thompson(1996)}]{hardy1996likelihood}
Hardy~RJ and Thompson~SG.
A likelihood approach to meta-analysis with random effects.
\textit{Stat. Med.} 1996; \textbf{15}: 619--629.

\bibitem[{Huizenga et~al.(2011)Huizenga et al.}]{huizenga2011testing}
Huizenga~HM, Visser~i and Dolan~CV.
Testing overall and moderator effects in random effects meta-regression.
\textit{Br. J. Math. Stat. Psychol.} 2011; \textbf{64}: 1--19.

\bibitem[{Jackson \& Bowden(2009)}]{jackson2009re}
Jackson~D and Bowden~J.
A re-evaluation of the ``quantile approximation method" for random effects meta-analysis.
\textit{Stat. Med.} 2009; \textbf{28}: 338--348.

\bibitem[{Kenne~Pagui et al.(2016)Kenne Pagui et al.}]{pagui2016median}
Kenne Pagui~EC, Salvan~A and Sartori~N.
Median bias reduction of maximum likelihood estimates.
\textit{Biometrika} 2017; \textbf{104}: 923--938.

\bibitem[{Knapp \& Hartung(2003)}]{knapp2003improved}
Knapp~G and Hartung~J.
Improved tests for a random effects meta-regression with a single covariate.
\textit{Stat. Med.} 2003; \textbf{22}: 2693--2710.

\bibitem[{Kosmidis \& Firth(2009)}]{kosmidis2009bias}
Kosmidis~I and Firth~D.
Bias reduction in exponential family nonlinear models.
\textit{Biometrika} 2009; \textbf{96}: 793--804.

\bibitem[{Kosmidis et~al.(2017)Kosmidis et al.}]{kosmidis2017improving}
Kosmidis~I, Guolo~A and Varin~C.
Improving the accuracy of likelihood-based inference in meta-analysis and meta-regression.
\textit{Biometrika} 2017; \textbf{104}: 489--496.

\bibitem[{Larsson \& Orsini(2014)}]{larsson2014red}
Larsson~SC and Orsini~N.
Red meat and processed meat consumption and all-cause mortality: A meta-analysis.
\textit{Am. J. Epidemiol.} 2014; \textbf{179}: 282--289.

\bibitem[{Pace \& Salvan(1997)}]{pace1997principles}
Pace~L and Salvan~A.
\textit{Principles of statistical inference: From a neo-Fisherian perspective}. 1997; London: World Scientific.

\bibitem[{Taubert et~al.(2007)Taubert, Roesen \&
  Sch{\"o}mig}]{taubert2007effect}
Taubert~D, Roesen~R and Sch{\"o}mig~E.
Effect of cocoa and tea intake on blood pressure: a meta-analysis.
\textit{Arch. Intern. Med.} 2007; \textbf{167}: 626--634.

\bibitem[{Van~Houwelingen et~al.(2002)Van~Houwelingen, Arends \& Stijnen}]{van2002advanced}
Van~Houwelingen~HC, Arends~LR and Stijnen~T.
Advanced methods in meta-analysis: multivariate approach and meta-regression.
\textit{Stat. Med.} 2002; \textbf{21}: 589--624.

\bibitem[{Zeng \& Lin(2015)}]{zeng2015random}
Zeng~D and Lin~DY.
On random-effects meta-analysis.
\textit{Biometrika} 2015; \textbf{102}: 281--294.

\end{thebibliography}
\end{document}